\documentclass[fleqn,usenatbib]{mnras}
\usepackage{newtxtext,newtxmath}
\usepackage[T1]{fontenc}
\usepackage{ae,aecompl}
\usepackage{graphicx}	
\usepackage{amsmath}	
\usepackage{color}

\newcommand{\hGpc}{$ \, h^{-1} \rm Gpc$}
\newcommand{\hMpc}{$ \, h^{-1}  \rm Mpc$}
\newcommand{\hMpcinv}{$ \, h \, \rm Mpc^{-1}$}
\newcommand{\hMpccube}{$(h^{-1} \rm Mpc)^3$}
\newcommand{\hkpc}{$ \, h^{-1} \rm kpc$}

\newcommand{\hMsun}{$\, h^{-1} \rm M_{\odot}$}
\newcommand{\Msun}{$\rm M_{\odot}$}

\newcommand{\nugc}{$\nu^2$GC}
\newcommand{\lsim}{\mbox{${\,\hbox{\hbox{$ < $}\kern -0.8em \lower 1.0ex\hbox{$\sim$}}\,}$}}
\newcommand{\gsim}{\mbox{${\,\hbox{\hbox{$ > $}\kern -0.8em \lower 1.0ex\hbox{$\sim$}}\,}$}}

\newcommand{\lensepaper}{(Metcalf et al. in preparation)}
\newcommand{\urluchuu}{\url{http://skiesanduniverses.org/}}
\newcommand{\urltao}{\url{https://tao.asvo.org.au/}}
\newcommand{\SU}{{Skies \& Universes}}

\title[The Uchuu simulations: Data Release 1]{The Uchuu Simulations: Data Release 1 and Dark Matter Halo Concentrations}

\author[Ishiyama et al.]{
Tomoaki Ishiyama,$^{1}$\thanks{E-mail: ishiyama@chiba-u.jp}
Francisco Prada,$^{2}$
Anatoly A. Klypin,$^{3,4}$
Manodeep Sinha,$^{5,6}$
\newauthor{
R. Benton Metcalf,$^{7}$
Eric Jullo,$^{8}$
Bruno Altieri,$^{9}$
Sof\'ia A. Cora,$^{10,11}$
Darren Croton,$^{5,6}$
}
\newauthor{
Sylvain de la Torre, $^{8}$
David E. Mill\'an-Calero,$^{2}$
Taira Oogi,$^{1}$
Jos\'e Ruedas,$^{2}$
}
\newauthor{
Cristian A. Vega-Mart\'inez$^{12,13}$
}
\\
$^{1}$Institute of Management and Information Technologies, Chiba University, 1-33, Yayoi-cho, Inage-ku, Chiba, 263-8522, Japan \\
$^{2}$Instituto de Astrof\'isica de Andaluc\'ia (CSIC), Glorieta de la Astronom\'ia, E-18080 Granada, Spain \\
$^{3}$Astronomy Department, New Mexico State University, Las Cruces, NM, USA \\
$^{4}$Department of Astronomy, University of Virginia, Charlettesville, VA, USA \\
$^{5}$Centre for Astrophysics \& Supercomputing, Swinburne University of Technology, 1 Alfred St., Hawthorn, VIC 3122, Australia \\
$^{6}$ARC Centre of Excellence for All Sky Astrophysics in 3 Dimensions (ASTRO 3D) \\
$^{7}$Dipartimento di Fisica \& Astronomia, Universit\`a di Bologna, via Gobetti 93/2, 40129 Bologna, Italy \\
$^{8}$ Aix Marseille Univ, CNRS, CNES, LAM, F-13388 Marseille, France \\
$^{9}$ European Space Astronomy Centre, ESA, Villanueva de la Ca\~nada,  E-28691 Madrid, Spain \\
$^{10}$ Instituto de Astrof\'isica de La Plata (CCT La Plata, CONICET, UNLP), Observatorio Astron\'omico,\\
Paseo del Bosque, B1900FWA La Plata, Argentina \\
$^{11}$ Facultad de Ciencias Astron\'omicas y Geof\'isicas, Universidad Nacional de La Plata, Observatorio Astron\'omico\\
Paseo del Bosque, B1900FWA La Plata, Argentina \\
$^{12}$ Instituto de Investigaci\'on Multidisciplinar en Ciencia y Tecnolog\'ia, Universidad de La Serena, Ra\'ul Bitr\'an 1305, La Serena, Chile\\
$^{13}$ Departamento de Astronom\'ia, Universidad de La Serena, Av. Juan Cisternas 1200 Norte, La Serena, Chile
}

\date{Accepted XXX. Received YYY; in original form ZZZ}
\pubyear{XXX}

\begin{document}
\label{firstpage}
\pagerange{\pageref{firstpage}--\pageref{lastpage}}
\maketitle 

\begin{abstract} 

We introduce the Uchuu suite of large high-resolution cosmological $N$-body simulations.  
The largest simulation, named Uchuu,
consists of 2.1 trillion ($12800^3$) dark matter particles in a box of side-length 2.0\hGpc, 
with particle mass $3.27 \times 10^{8}$\hMsun.
The highest resolution simulation, Shin-Uchuu, consists of 262 billion ($6400^3$)
particles in a box of side-length 140\hMpc, with particle mass $8.97 \times 10^{5}$\hMsun.
Combining these simulations we can follow the evolution of dark matter halos and subhalos 
spanning those hosting dwarf galaxies to massive galaxy clusters across an unprecedented volume. 
In this first paper, we present basic statistics, dark matter power spectra, 
and the halo and subhalo mass functions, 
which demonstrate the wide dynamic range and superb statistics of the Uchuu suite.
From an analysis of the evolution of the power spectra we conclude that
our simulations remain accurate from the Baryon Acoustic
Oscillation scale down to the very small.  We also provide parameters of
a mass-concentration model, which describes the evolution of
halo concentration and reproduces our simulation data to within 5 per cent
for halos with masses spanning nearly eight orders of magnitude at redshift $0 \leq z \leq 14$.
There is an upturn in the mass-concentration relation for the population 
of all halos and of relaxed halos at $z\gtrsim 0.5$, whereas no upturn is detected at $z<0.5$.
We make publicly available various $N$-body products as part of Uchuu Data Release 1 
on the \SU\ site\footnote{http://skiesanduniverses.org/}. 
Future releases will include gravitational lensing maps 
and mock galaxy, X-ray cluster, and active galactic nuclei catalogues.
\end{abstract}


\section{Introduction}\label{sec:intro}

Over the last three decades, our knowledge of the
formation of structure in the Universe and its growth has been dramatically advancing.
The lambda cold dark matter ($\Lambda$CDM) paradigm is now regarded as the standard model
to describe our Universe, in which structure formation occurs
hierarchically.  Smaller scale structures gravitationally collapse
first and form dark matter halos, which repeats with time
to produce larger and larger scale structures. 

Luminous objects, such as galaxies and active galactic nuclei (AGN),
form in the centres of halos and trace this growth in structure. 
From their light we can extract valuable information 
about the underlying cosmology.  The properties of galaxies and
AGN at various epochs also contain information that reveals the 
physics of their formation.  Ongoing/upcoming wide and deep
photometric/spectroscopic surveys performed by the 
Subaru Hyper Suprime-Cam (HSC: \citealt{Miyazaki2006, Miyazaki2012}), 
the Subaru Prime Focus Spectrograph (PFS: \citealt{Takada2014}), 
and Euclid \citep{Euclid2011}, amongst others, will offer vast amounts of data on galaxies and AGN, 
covering many Gpc in scale. 
This data revolution is driving our understand the Universe to unprecedented detail.
However to be successful, such experiments require percent-level precision theoretical predictions
in order to quantify potential deviations from $\Lambda$CDM, to separate out 
cosmological from baryonic effects, and to identify other sources of systematic error.
 
To extract as much information as possible from such observations,
we need detailed ``mock'' galaxy and AGN catalogues. Traditionally, dark matter only
cosmological $N$-body simulations have played a central role in
this effort, especially when coupled with semi-analytic galaxy formation models \citep[e.g.,
][]{Benson2012, Croton2016, Makiya2016, Cora2018, Shirakata2019}.  
Several catalogs and simulations of this sort are publicly available. 
These include the Millennium simulation galaxy catalogues \citep{Springel2005, Croton2006}, 
galaxies built on the MultiDark simulation \citep{Prada2012, Klypin2016, Knebe2018}, 
as well as \nugc\ \citep{Ishiyama2015, Makiya2016} 
(see, for example, \urluchuu and \urltao). 

More recently, galaxy catalogues using cosmological hydrodynamical simulations 
have become available, such as EAGLE
\citep{Schaye2015}, Horizon-AGN \citep{Dubois2014}, Illustris \citep{Vogelsberger2014,
  Springel2018}, and Magneticum \citep{Dolag2015} (for a recent review see \citealt{Vogelsberger2020}). 
However, their current simulation size is
typically of the order $\sim 100$\hMpc\ in scale, which is much smaller than the
observational surveys looking to extract cosmological information. 
Therefore, large $N$-body simulations and semi-analytic galaxy formation models
remain an essential way to 
construct appropriate mock catalogues for comparison to the observations.

The challenge lies in producing the most closely matched mock catalogues as possible.
For this, the simulation volume must 
be comparable with those covered by the surveys, and the mass resolution must be sufficient
to resolve those galaxies formed in the
early Universe that are expected to ultimately feature in the observations.  
This balance of size and resolution is critical. 
Nowadays, cosmological $N$-body simulations with more
than $10^{11}$ particles have been achieved on large modern supercomputers
\citep[e.g.,][]{Skillman2014, Ishiyama2015, Potter2017, Cheng2020}.  However,
none of them yet satisfy both requirements simultaneously.  For example, the 
Euclid Flagship simulation \citep{Potter2017} contains two trillion particles over a 
cosmologically large volume. But to achieve such a volume its mass resolution is
$\sim 10^{9}$\hMsun, a high mass limit in the early Universe where the need
is to accurately resolve the progenitor galaxies that will be observed.

There are further complications. 
Although simulations like the Euclid Flagship simulation leverage the
latest supercomputing hardware and algorithms to get ever so closer
to producing an ideal mock universe, linking the evolution of
the unique structures within across time remains a significant hurdle. 
This linking takes the form of 
halo/subhalo merger trees, which are required by modern semi-analytic
galaxy models to construct mock catalogues. Without these, simulation teams must rely on
empirical models to construct their mock universes, like abundance matching 
\citep[e.g.,][]{Kravtsov2004c, Vale2004, Conroy2006, Shanker2006, Moster2010, Behroozi2010, Reddick2013} 
and the halo occupation distribution method 
\citep[e.g.,][]{Peacock2000, Seljak2000, Berlind2002, Cooray2002, Zheng2007, Leauthaud2012}. 
Such mocks thus lack evolutionary information for individual objects (halos or galaxies),
and their use for analysis and interpretation of the observations can be limited. 
Also, given their statistical nature, empirical models are further removed from
the physics of galaxy formation when compared to semi-analytic galaxy formation models.

To address these problems, we have run a suite of ultra-large
cosmological {\it N}-body simulations, the 
Uchuu\footnote{Uchuu means ``Universe'' in Japanese.}
simulation suite, on which the halo/subhalo merger trees have been constructed.
The largest Uchuu simulation consists of $12800^3\approx 2.1$~trillion dark matter particles
in a box of 2.0\hGpc\ side-length; this results in a mass of each dark matter particle of
$3.27 \times 10^{8}$\hMsun. 
The highest resolution simulation in the suite 
consists of $6400^3$ particles in a 140\hMpc\ box, 
with particle mass $8.97 \times 10^{5}$\hMsun. 
The cosmological parameters for all Uchuu simulations 
are based on the latest observational results
obtained by the Planck satellite \citep{Planck2020}. Thanks to its large
volume and high mass resolution, our simulation suite provides the most
accurate theoretical templates to construct galaxy and AGN catalogues to date.

In this paper, we introduce the Uchuu simulation suite and describe the public release
of various $N$-body dark matter data products. 
These include subsets particles drawn from each simulation,
matter power spectra, halo/subhalo catalogues, their merger trees, and
various lensing products. These comprise data release 1 (DR1).  Details of the lensing
products are given in a companion paper \lensepaper.  
Mock galaxy catalogs, constructed using three semi-analytic
models -- \nugc\ \citep{Makiya2016, Shirakata2019}, SAG \citep{Cora2018}, 
and SAGE \citep{Croton2016} -- will be released as an
official data release 2 (DR2) in the near future.
All data will be made publicly available on the \SU\ site at \urluchuu. 

The current paper is organised as follows: 
In \S \ref{sec:method}, we describe the simulation and numerical
methods used to produce the Uchuu data.  In \S \ref{sec:result}, we present matter power spectra and
mass functions as standard validations of our simulations, along with other
statistics measured at the unprecedented detail Uchuu affords. 
To conclude, our work is summarised in \S \ref{sec:summary}.

\section{Initial Conditions and Numerical Method}\label{sec:method}

The basic properties of the cosmological $N$-body simulations that 
comprise the Uchuu suite are listed in
Table~\ref{tab:sim}. The largest simulation, simply named Uchuu, was run using
$12800^3$ dark matter particles in a comoving box of side-length
2.0\hGpc, resulting in a dark matter particle mass resolution of $3.27 \times
10^{8}$\hMsun. The gravitational softening length of Uchuu is 4.27\hkpc.  We
additionally produced two smaller volume simulations 
that are otherwise identical to the full Uchuu simulation. 
The first, named mini-Uchuu, was run with
$2560^3$ particles in a 400\hMpc\ comoving side-length box. The second, named
micro-Uchuu, was run with $640^3$ particles in a 100\hMpc\ comoving side-length box.
These two simulations have a considerably smaller data footprint, and hence act as 
convenient testing simulations, amongst other applications. Finally, to push the 
resolution of Uchuu to lower mass halos, a higher-resolution simulation was run, named
Shin-Uchuu\footnote{Shin means ``deep'' in Japanese.}, using
$6400^3$ particles. This simulation has a comoving box of side-length 140\hMpc\ and
particle mass of $8.97 \times 10^{5}$\hMsun. Its
gravitational softening length is 0.4\hkpc.

\begin{table}
\centering
\caption{
Properties of the cosmological $N$-body simulation suite presented in this work. 
Here, $N$, $L$, $\varepsilon$, and $m_{\rm p}$ are the total number of particles, box
length, softening length, and particle mass resolution, respectively.
The first four rows are for the Uchuu suite. 
The next is the Phi-4096 simulation used to parameterise the mass-concentration relation, as described in \S\ref{sec:m-c}.
Lastly is the \textsc{glam} simulation used to evaluate cosmic variance on the power spectrum,
 the details of which are given in \S\ref{sec:pk}.
}
\label{tab:sim}
\begin{tabular}{lccccl}
\hline
\shortstack{Name \\ {}}  & \shortstack{$N$ \\ {}} & \shortstack{$L$ \\(\hMpc)} &  \shortstack{$\varepsilon$ \\ (\hkpc)} &  \shortstack{$m_{\rm p}$ \\ (\hMsun)}\\
\hline
Uchuu & $12800^3$ & 2000.0 & $4.27$ & $3.27 \times 10^{8}$ \\
mini-Uchuu & $2560^3$ & 400.0 & $4.27$ & $3.27 \times 10^{8}$ \\
micro-Uchuu & $640^3$ & 100.0 & $4.27$ & $3.27 \times 10^{8}$ \\
Shin-Uchuu & $6400^3$ & 140.0 & $0.4$ & $8.97 \times 10^{5}$ \\
\hline 
Phi-4096 & $4096^3$ & 16.0 & $0.06$ & $5.13 \times 10^{3}$ \\
\hline 
\textsc{glam} & $2000^3$ & 2000.0 & $ 286 $ & $8.60 \times 10^{10}$ \\
\hline 
\end{tabular}
\end{table}

\begin{figure*}
\centering 
\includegraphics[width=16cm]{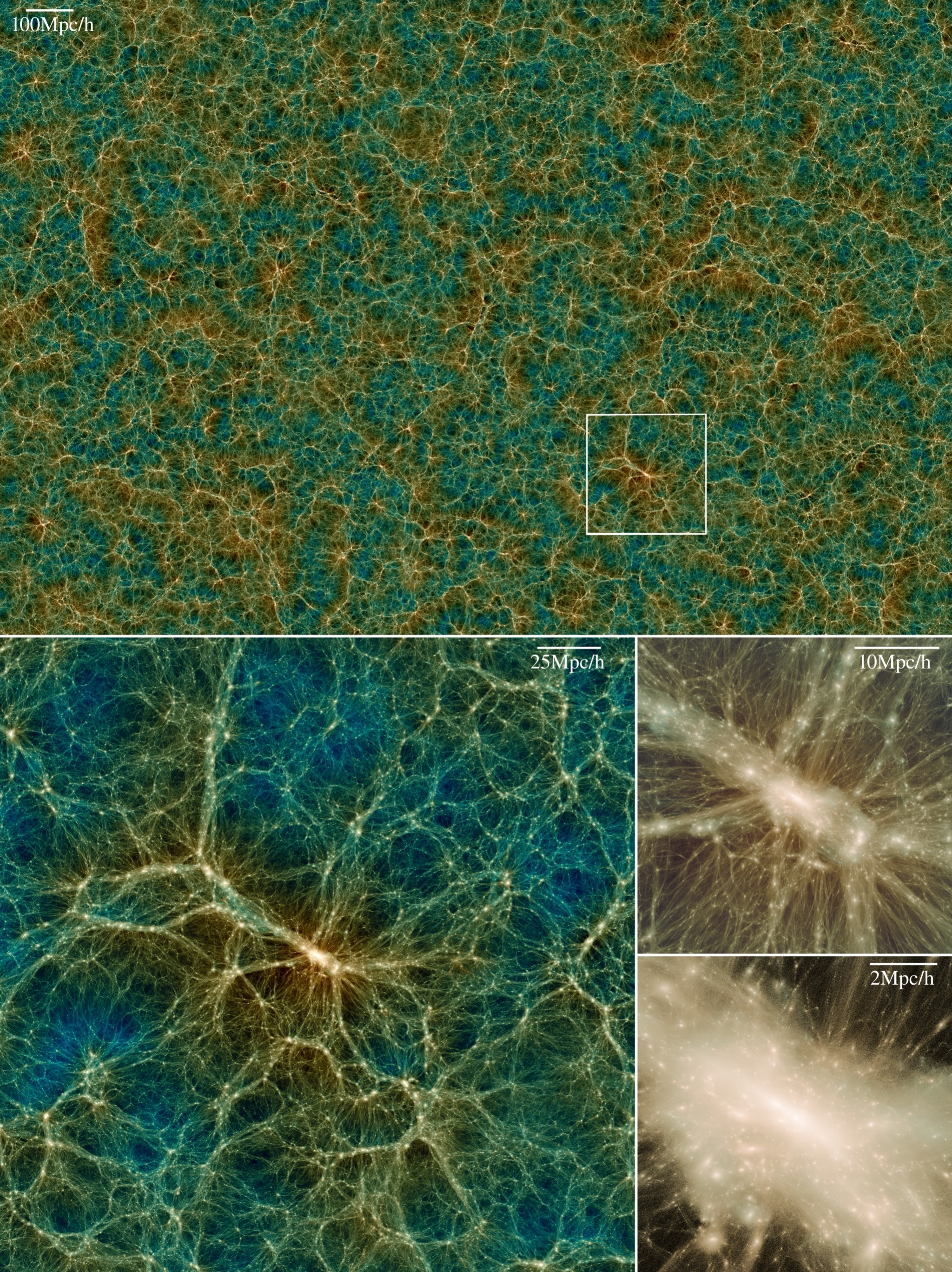} 
\caption{ 
Dark matter distribution of the Uchuu simulation at $z=0$. 
The top panel shows a 2000 \hMpc\ $\times$ 1333 \hMpc\ slice of thickness of 25 \hMpc.  
The three bottom panels zoom in on largest halo in Uchuu, on scales of 
250, 38, and 9.4 \hMpc\ (moving clockwise). 
The white box in the top panel is the same region visualized in the left-bottom panel, 
equivalent in scale to the Bolshoi simulation \citep{Klypin2011}.
The evolution of the largest halo is visualized in Figure~\ref{fig:snapshot3}.
}
\label{fig:snapshot1}
\end{figure*}

\begin{figure*}
\centering 
\includegraphics[width=16cm]{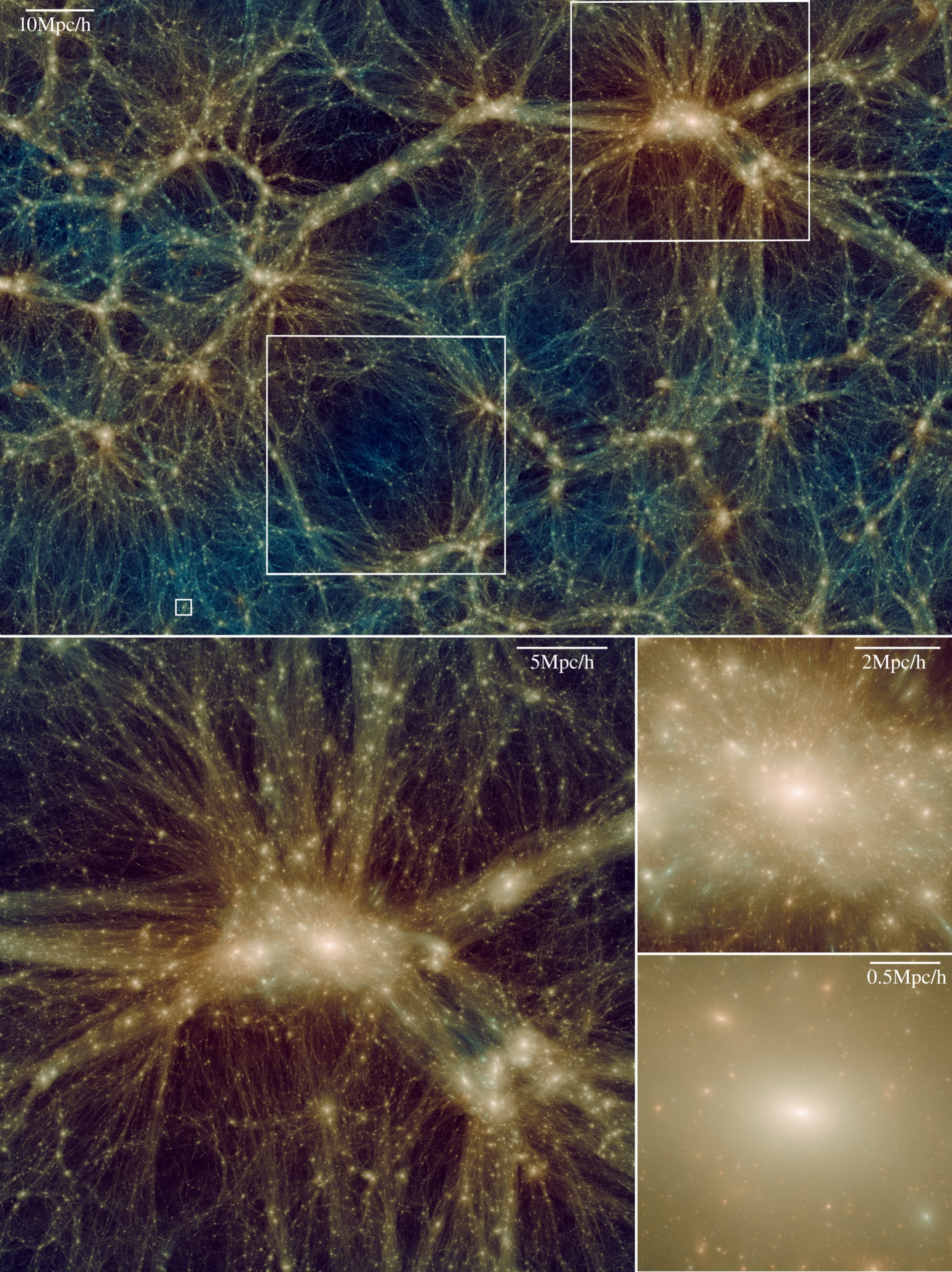} 
\caption{ 
Dark matter distribution of the higher resolution Shin-Uchuu simulation at $z=0$. 
The top panel shows a 140 \hMpc\ $\times$ 93 \hMpc\ slice of thickness of 17.5 \hMpc.  
The three bottom panels zoom in on the largest halo in Shin-Uchuu, on scales of 
35, 7.5, and 2.3 \hMpc\ (moving clockwise). 
The upper-most white box in the top panel is the same region visualized in the left-bottom panel.
The middle white box is a void region selected by eye, 
while the lower-most white box is a Milky-way sized halo. 
The evolutionary paths of the latter two regions are visualized in Figure~\ref{fig:snapshot3}.
}
\label{fig:snapshot2}
\end{figure*}

\begin{figure*}
\centering 
\includegraphics[width=16cm]{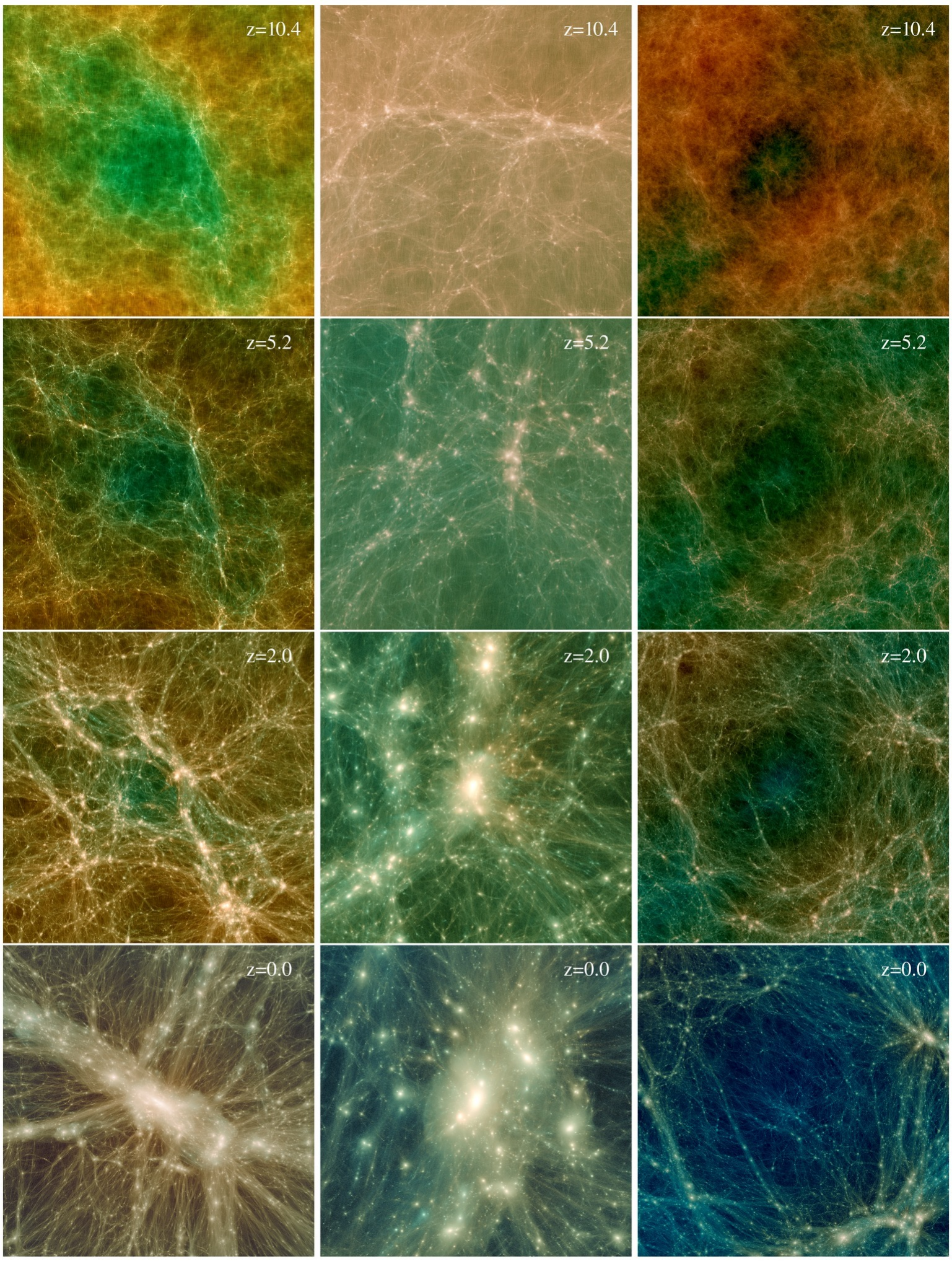} 
\caption{ 
Evolution of three different environments drawn from the Uchuu simulation suite. 
Redshift is marked in each panel, $z=10.4$, 5.2, 2.0, and 0.0, from top to bottom. 
The three environments are: left, the largest halo in the Uchuu simulation; 
middle, a Milky Way sized halo in the Shin-Uchuu simulation; and 
right, a void selected by eye in the Shin-Uchuu simulation. 
The side lengths shown are 38, 2.3, and 35 \hMpc\ for the left, middle, and right panels, respectively.
}
\label{fig:snapshot3}
\end{figure*}

To generate the initial conditions for each simulation, we used the parallel
\textsc{2LPTic} code\footnote{\url{http://cosmo.nyu.edu/roman/2LPT/}}
adopting second-order Lagrangian perturbation theory
\citep[][]{Crocce2006}.  
We calculated the matter transfer function using the online
version of \textsc{Camb}\footnote{\url{http://lambda.gsfc.nasa.gov/toolbox/tb\_camb\_form.cfm}}
\citep{Lewis2000}.  
Throughout, the latest cosmological parameters measured by 
the Planck Satellite \citep{Planck2020} were adopted: $\Omega_0=0.3089$,
$\Omega_{\rm b}=0.0486$, $\lambda_0=0.6911$, $h=0.6774$, $n_{\rm s}=0.9667$, and
$\sigma_8=0.8159$. The initial redshift for all simulations was 127.
The gravitational evolution of each dark matter particle was calculated using the massively parallel
TreePM code, \textsc{GreeM}\footnote{\url{http://hpc.imit.chiba-u.jp/~ishiymtm/greem/}}
\citep{Ishiyama2009b, Ishiyama2012}. This was run
on the Aterui II supercomputer at the Center for Computational Astrophysics (CfCA), 
National Astronomical Observatory of Japan, and the K
computer at the RIKEN Advanced Institute for Computational Science.
The evaluation of the tree forces was accelerated by the 
\textsc{Phantom-grape}\footnote{\url{https://bitbucket.org/kohji/phantom-grape/}} software
\citep{Nitadori2006, Tanikawa2012, Tanikawa2013, Yoshikawa2018}.
50 particle snapshots covering $z=14$ to 0 were written to disk. 

To identify bound structures within the particle data at each snapshot, the 
\textsc{rockstar}\footnote{\url{https://bitbucket.org/gfcstanford/rockstar/}}
phase space halo/subhalo finder \citep{Behroozi2013} was run. We then 
constructed merger trees using the 
\textsc{consistent trees}\footnote{\url{https://bitbucket.org/pbehroozi/consistent-trees/}}
code \citep{Behroozi2013b}. However, by default, \textsc{consistent trees} does not run in 
parallel in a distributed memory environment. Hence, our use of \textsc{consistent trees} 
needed to be modified to make merger tree construction possible on simulations of the
size of those in the Uchuu suite.

To overcome this hurdle, we split the full box of each simulation
into smaller regularly spaced sub-volumes, and ran \textsc{consistent trees} for each
sub-volume separately. In the case of the Uchuu
simulation, the \textsc{rockstar} catalogues for each snapshot were broken 
into $20^3$ sub-catalogues of comoving volume $2000^3 / 20^3 =
100^3$ \hMpccube. These sub-catalogues were then grouped across
redshift to construct the merger trees, where sub-catalogues in
each group occupy the same Euler comoving volume.  However, halos and
subhalos can have progenitor/descendant relationships that lie outside of such defined 
groups.  To account for this, we added halos and
subhalos to the group if they lay within a 25
\hMpc\ distance from any edge of the corresponding Euler volume of
the group.  However, this also meant that added halos and subhalos could be duplicated in multiple
groups, and after running \textsc{consistent trees} across all groups, multiple
merger tree files constructed separately could have the same trees that
contain these overlapping halos and subhalos.  To correct for this, we ran a final step to clean 
such overlapping trees from all tree files except for one, thus removing any duplicates.
In this way the merger trees are consistently and efficiently constructed for the entire box.
To show the robustness of this method, in Appendix \ref{sec:merger_tree}
we compare merger trees from the mini-Uchuu simulation constructed by 
this method and the original \textsc{consistent trees} code.

In addition to the Uchuu and Shin-Uchuu simulations, we
use an additional small volume but ultra-high resolution simulation, Phi-4096, 
described in Table~\ref{tab:sim}, focused on $z>7.5$. 
This simulation extends the halo mass range down to $10^7$\Msun\ at
high redshift and helps with the evaluation of resolution effects. Phi-4096
is also used when we later examine the halo mass-concentration relation. 
The cosmological parameters of Phi-4096 are slightly different from
those of the Uchuu suite: 1.6 per cent larger for $\sigma_8$ 
and less than 1 per cent for the remaining parameters. 
\citet{Dutton2014} show that the average 
concentration of a halo increases by about 8 per cent when $\sigma_8$  
increases by  7.9 per cent. The latter is much larger than the 
difference between the Uchuu and the Phi-4096 simulations, 
and thus we treat all simulations equally in what follows.
The initial conditions of Phi-4096 were generated by the publicly
available code, \textsc{music}\footnote{\url{https://bitbucket.org/ohahn/music/}}
\citep{Hahn2011}.
All other numerical tools used to process and study Phi-4096 were the same as
those for the Uchuu simulations. 

Figure~\ref{fig:snapshot1} shows dark matter distribution of the Uchuu
simulation at $z=0$.  The top panel shows a 2000 \hMpc\ $\times$ 1333
\hMpc\ slice of thickness of 25 \hMpc. 
To calculate density fields, we used the tetrahedron quadrupole particle 
mesh (T4PM) method \citep{Hahn2013}.
Vast cosmic web structures are clearly visible.  The three bottom panels are close-ups of
the largest halo in the box at different zoom levels, highlighting the detail Uchuu can resolve.  
The mass of this 
halo is $\sim 5.0 \times 10^{15}$\hMsun.  More than a thousand halos
with mass greater than $\sim 1.0 \times 10^{15}$\hMsun\ exist in
Uchuu, which allows us to study such rare
objects with unprecedented statistics.  Each such halo is resolved by
more than three million particles.

Figure~\ref{fig:snapshot2} shows dark matter distribution of the Shin-Uchuu
simulation at $z=0$.  The top panel shows a 140 \hMpc\ $\times$ 93
\hMpc\ slice of thickness of 17.5 \hMpc. 
The format is the same as in Figure~\ref{fig:snapshot1},
where we again zoom in on the largest structure in the box. 
Figure~\ref{fig:snapshot3} shows the evolution of 
three selected regions at redshifts $z=10.4$, 5.2, 2.0, and 0.0: 
the largest cluster halo in the Uchuu simulation, a Milky Way sized halo in the Shin-Uchuu simulation, 
and a void selected by eye in the Shin-Uchuu simulation.
The evolution of the cosmic web leading to protocluster formation is clearly 
visible in the left row. 
In the center row, thanks to the high resolution of the Shin-Uchuu simulation 
we observe many progenitor halos of the Milky-Way system, even at $z>10$.
More than ten thousand of Milky-Way sized halos are formed in Shin-Uchuu, 
which enables us to study the statistics of such halos with unprecedented resolution. 
In the right row, which tracks the evolution of a large cosmic void, 
we see how small halos form even within such an underdense environment, 
highlighting both the large spatial volume and the high resolution of Shin-Uchuu. 

\begin{figure}
\centering 
\includegraphics[width=8.4cm]{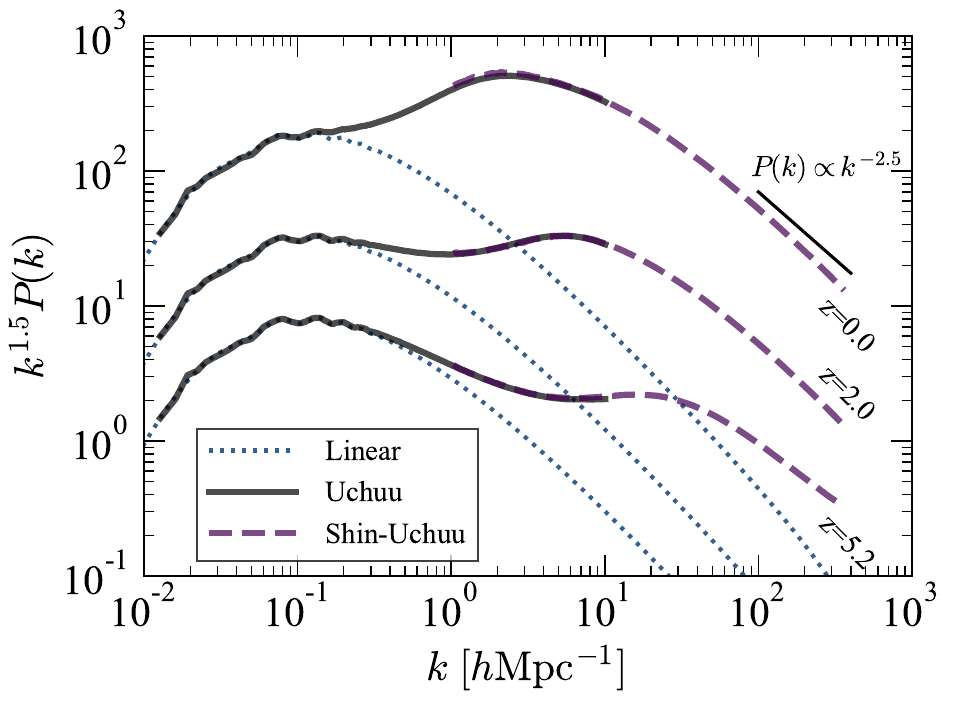} 
\caption{
Power spectra of the Uchuu (solid curves) and Shin-Uchuu (dashed curves)
simulations. Results are shown for $z=0.0, 2.0$, and 5.2 (top to bottom), 
where each are multiplied by $k^{1.5}$ 
to more clearly highlight the BAO features. 
The dotted curves show the linear power spectra. 
At high-$k$, the power is close to $P(k) \propto k^{-2.5}$.
}
\label{fig:pk}
\end{figure}

\begin{figure*}
\centering 
\includegraphics[width=16cm]{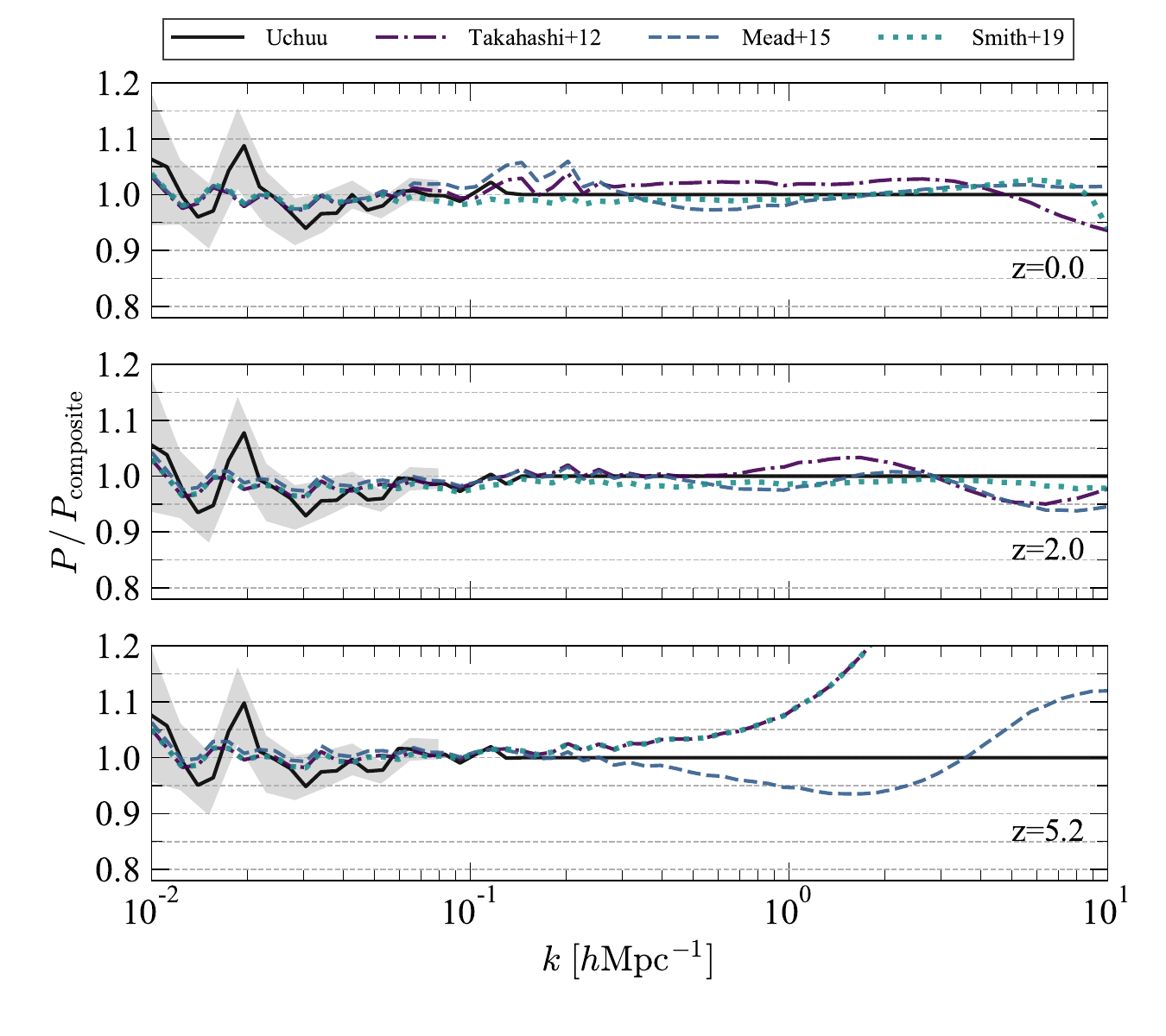} 
\caption{
Comparison of power spectra at low-$k$
with predictions from three models of 
the non-linear matter power spectrum 
\citep{Takahashi2012, Mead2015, Smith2019}. 
The top, middle, and bottom panels show 
the results for $z=0.0, 2.0$, and $5.2$, respectively.  
The reference spectra, $P_{\rm composite}$, 
are taken from the Uchuu simulation
for $k>0.1$ \hMpcinv\ and an ensemble of \textsc{glam} simulations
\citep{Klypin2018} for $k<0.1$ \hMpcinv.
The grey shaded regions are the estimated $1\sigma$ scatter of 
the Uchuu simulation.
}
\label{fig:pk2}
\end{figure*}

\section{Results}\label{sec:result}

We first present the basic properties of halos, matter
power spectra, and mass functions of the Uchuu simulations.
Then we propose an accurate model of halo concentrations that describes our simulations.

\subsection{Power spectrum}\label{sec:pk}

Large number of particles and high force resolution of Uchuu simulations allow us to measure the power spectrum of dark matter fluctuations for a very large range of wavenumbers. However, in order to use this potential advantage, we need to implement a technique that is different from traditionally used method based on a large single mesh. We start with a single mesh of size $N_{\rm g}=2400$ in one dimension. The mesh covers the whole computational box $L$. The Cloud-In-Cell technique is applied to all particles $N_{\rm p}$ to find the dark matter density on this mesh. In this process the "weight" of each particle is assumed to be $W_1= N_{\rm g}^3/N_{\rm p}$. In this case the total "weight" of all particles is simply the total number of grid points. Using the FFT we find the power spectrum of the density field. Because the mesh is not very large, it lacks information on wavenumbers larger than the Nyquist frequency $k_{\rm Ny}= (\pi/L)N_{\rm g}$ of the mesh.

On the next step we again use the $N_{\rm g}^3$ mesh, but this time it covers only half of the computational box, and, thus, it has box size $L/2$, and its Nyquist frequency is  twice larger. When estimating the density,  the coordinates of each particle are folded with the new box size: $\tilde x = \mod(x,L/2)$ and the same is applied for $y-$ and $z-$coordinates. The density assignment is made using the new coordinates $(\tilde x,\tilde y,\tilde z)$ and the particle weight is increased by factor $2^{3/2}$. Again, we use the FFT to find the power spectrum. This time it does not have any information about waves longer than $L/2$, but it has information on frequencies that are twice higher than on the first step. 
We proceed with new levels each time reducing twice the box size, increasing twice the mesh Nyquist frequency and increasing particle "weight" by $2^{3/2}$. In total we have six levels of meshes with the combined range of wavenumbers from $(2\pi/L)$ to $2^5 N_{\rm g}(2\pi/L)$. 

The final power spectrum of the whole simulation is constructed from the set of six power spectra obtained for each mesh. We start with the first power spectrum corresponding to the box $L$. 
The power spectrum is corrected for the aliasing. All estimates are taken until we reach the points where the aliasing correction falls below $0.85$. At this moment we proceed to the results of the next mesh with twice smaller box size. The process continues for all six levels of meshes.

Figure~\ref{fig:pk} shows the evolution of the power spectra of
dark matter from the Uchuu and Shin-Uchuu simulations.  
The Uchuu suite allows us to plot a large range of wave numbers covering five orders of
magnitude. At low-$k$, the simulation follows the linear matter
power spectrum. Power spectra are multiplied by $k^{1.5}$, where $k$ is the wave number,
which clearly highlights the Baryon Acoustic Oscillations (BAO) features, including 
the first and second peaks at $k \sim 0.07$ and 0.13 \hMpcinv\ for $z<2$.
Other small peaks can be seen at $z>2$, but these are gradually smeared out at lower redshift as
the spectra is influenced by non-linear structure growth.
Both simulations agree well with each other in the overlapping
region, $1 < k < 10$ \hMpcinv.
The agreement is similar to that found by \citet{Schneider2016},
who compared power spectra calculated by three different $N$-body methods 
and showed convergence at the one percent level for $k<1$ \hMpcinv, and three percent level for
$k<10$ \hMpcinv when using standard run parameters.
In their tests, \textsc{gadget-3} \citep{Springel2005b} was adopted, and
its time-stepping method and choice of run parameters
are similar to what was used in the Uchuu and the Shin-Uchuu simulations.

Figure~\ref{fig:pk2} shows a comparison between the Uchuu matter power spectra at low to middle-$k$
with predictions from three models of nonlinear structure evolution
\citep{Takahashi2012, Mead2015, Smith2019}. The spectra of \citet{Mead2015} are calculated using
\textsc{hmcode}\footnote{\url{https://github.com/alexander-mead/hmcode}}, 
and of \citet{Takahashi2012} and \citet{Smith2019} using 
\textsc{ngenhalofit}\footnote{\url{https://CosmologyCode@bitbucket.org/ngenhalofitteam/ngenhalofitpublic.git}}.
We construct the reference spectra $P_{\rm composite}$ as a composite of the Uchuu power spectrum
for $k>0.1$ \hMpcinv, and an average of an ensemble of 17 \textsc{glam} Particle-Mesh simulations
\citep{Klypin2018} for $k<0.1$ \hMpcinv\ to ameliorate the effects of
cosmic variance. The \textsc{glam} simulations were run in $2h^{-1}{\rm Gpc}$ side-length boxes 
with $2000^3$ particles moving in a $7000^3$ mesh, providing $8.6\times 10^{10}$\hMsun\ 
mass per particle resolution.

The shaded regions in Figure~\ref{fig:pk2} show the estimated $1 \sigma$ scatter of the Uchuu
simulation, which overlaps with the composite power spectra, meaning
that deviations of the simulation spectra at $k<0.1$ \hMpcinv\ are
due to the cosmic variance. For $k>0.1$ \hMpcinv, the model
prediction of \citet{Smith2019} agrees remarkably well with the Uchuu
simulation at $z=0$ and $z=2$, where the error is within $\sim$
2 per cent up to $k=10$ \hMpcinv. As expected, this model fails to predict the
spectra at $z=5.2$ because it is not
calibrated beyond $z=3$ and reverts to that provided by
\citet{Takahashi2012}. The models of \citet{Takahashi2012} and \citet{Mead2015} predict
the power within $\sim$ 5 per cent at $z=0.0$ and $2.0$.
At $z=5.2$, \citet{Mead2015} predicts the power within $\sim$ 5 per cent at 
$k<4$ \hMpcinv, which is better than the others, 
even though it is not calibrated for such high redshifts.
In general, the accuracy at high redshift reflects how the underlying models are constructed.

These results demonstrate that our simulations are
accurate across a wide dynamic range, from BAO to very small
scales, reinforcing their usefulness for a wide range of applications.

\subsection{Halo mass function}\label{sec:massfunc}

\begin{figure}
\centering 
\includegraphics[width=9cm]{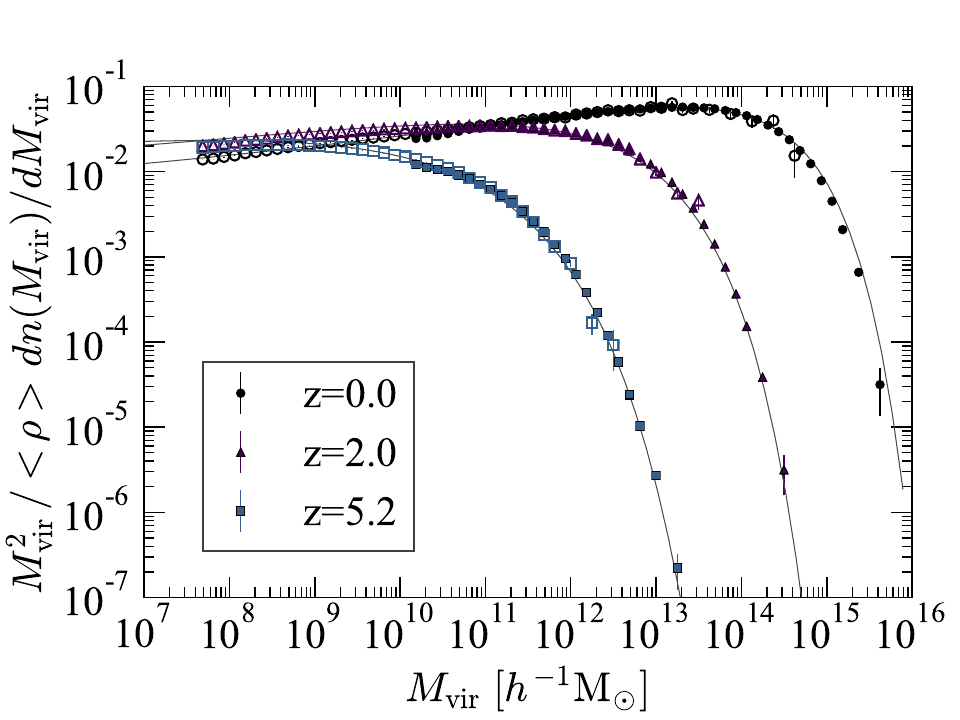} 
\caption{ 
Mass function of distinct halos 
multiplied by $M_{\rm vir}^2$ and divided by the mean cosmic density
from the Uchuu (filled symbols) and the Shin-Uchuu (open symbols)
simulations at $z=0.0$, $2.0$ and $5.2$. These cover approximately eight orders of magnitude
in halo mass. The error bars correspond to Poisson error. 
We only consider halos with more than 40 particles. 
Solid curves show the fitting functions proposed by \citet{Despali2016}. 
}
\label{fig:multiplicity}
\end{figure}

Figure~\ref{fig:multiplicity} shows the halo mass function multiplied by 
$M_{\rm vir}^2$ and divided by the mean cosmic density 
for the Uchuu and the Shin-Uchuu simulations at $z=0$, $2.0$, and $5.2$, 
where $M_{\rm vir}$ is the halo virial mass. 
We only consider halos with more than 40 particles, 
corresponding to a minimum mass of $1.3 \times 10^{10}$\hMsun\ for Uchuu and 
$3.6 \times 10^{7}$\hMsun\ for Shin-Uchuu. 
With this lower limit, the combined mass range spans
approximately eight orders of magnitude, reflecting the power of our
very large simulation set. The convergence between both simulations is
remarkably good, within a few per cent.

\begin{figure}
\centering 
\includegraphics[width=9cm]{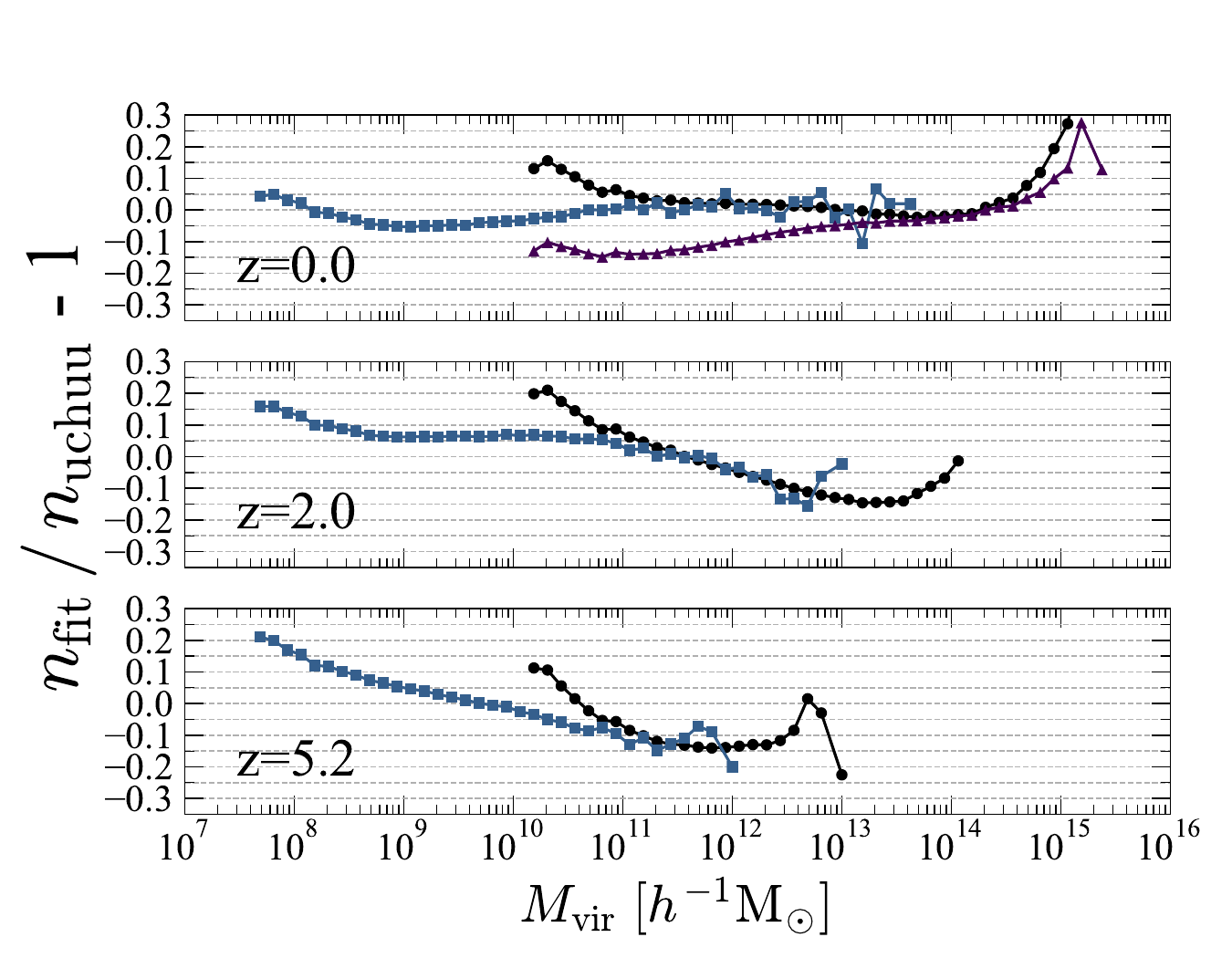} 
\caption{ 
Comparisons of the halo mass functions of Uchuu (filled circles) 
and Shin-Uchuu (filled squares) with the fitting functions proposed by \citet{Despali2016}. 
Triangles at $z=0$ compare against the specific high mass halo fitting 
function of \citep{Despali2016}, as described in the text. 
}
\label{fig:multiplicity_comp}
\end{figure}

Figure~\ref{fig:multiplicity} also considers the halo mass fitting function
proposed by \citet{Despali2016}, shown by each solid line. The fractional difference between
this function and each simulation is presented in 
Figure~\ref{fig:multiplicity_comp} for Uchuu (filled circles) and Shin-Uchuu (filled squares).
This comparison highlights a 5 per cent 
difference at $z=0$ 
over a broad range of halo mass up to $M_{\rm vir} < \sim 3 \times 10^{14}$\hMsun. 
\citet{Despali2016} present different parameterisations of their 
fitting function specifically for halos more massive than 
$M_{\rm vir} > 3 \times 10^{13}$\hMsun\ (shown in their Table~3).
These provide a better fit in our analysis for Uchuu halos of mass  
$M_{\rm vir} > 2 \times 10^{14}$\hMsun. 
Overall, we find a difference
of less than 10 per cent for halos with 
$M_{\rm vir} < 10^{15}$\hMsun. At $z=2.0$ and $5.2$, this difference increases to around 10 per cent. 
At the massive end of the mass function, the fitting function underestimates the number of
massive halos, while it overestimates the number of less massive halos.
We expect this is related to challenges of cosmic variance and the accuracy of the
simulations used to derive the fitting function (see also \citet{Knebe2013}).  

We do not perform an extensive analysis of the mass function in this paper, 
but leave it for a dedicated study at a later time.

\subsection{Subhalo mass function}\label{sec:subhalo_massfunc}

\begin{figure}
\centering 
\includegraphics[width=9cm]{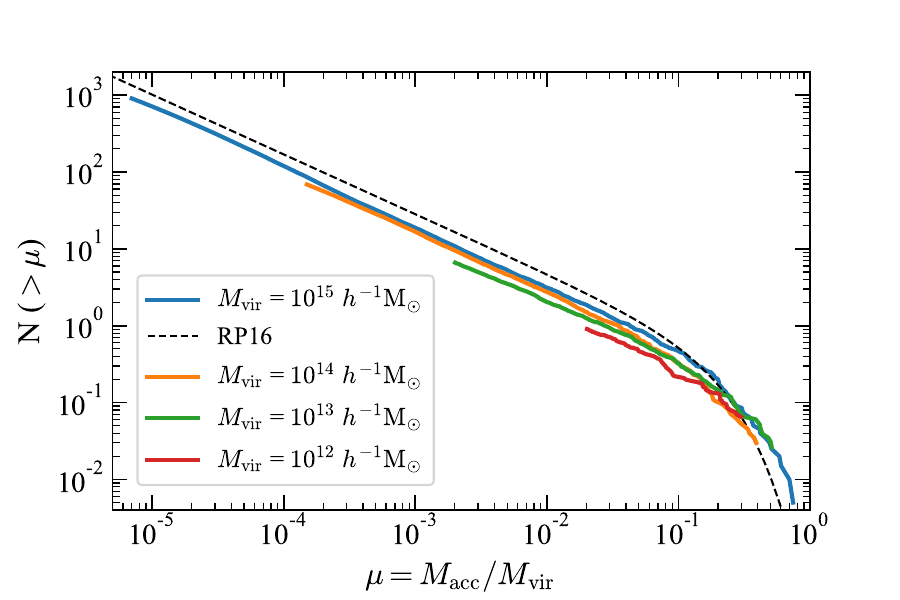} 
\caption{ 
The Uchuu subhalo mass function at $z=0$, $N(> \mu$), for different 
host halos as a function of $\mu = M_{\rm acc}/M_{\rm vir}$ (solid curves). 
As a reference, the dashed line shows the best-fitted Schechter-like function 
fit to the Bolshoi Planck / MultiDark Planck simulations \citep{2016MNRAS.462..893R}. 
Note the similar overall power-law slope but different shape at the exponential 
tail due to the much higher statistics of Uchuu.
}
\label{fig:subhalo_massfunction}
\end{figure}

In Figure~\ref{fig:subhalo_massfunction} we display the mean subhalo mass 
function for the Uchuu simulation at $z=0$, $N(> M_{\rm acc}/M_{\rm vir})$, 
computed in four narrow (5 per cent) bins of host halo mass,  
$M_{\rm vir} = 10^{15}, 10^{14}, 10^{13}, 10^{12}$\hMsun, 
where the subhalo mass $M_{\rm acc}$ is defined at the time of first accretion.
Solid lines with different colors correspond to different host halo mass bins as 
label in the figure panel. Overall our results are in good agreement with previous 
simulation works and model predictions 
\citep[e.g.,][and references therein]{2016MNRAS.462..893R, Hiroshima2018, 2018MNRAS.474.3043V,2018MNRAS.475.4066V} 
that explore the subhalo mass function shape and host halo mass dependence. 
This figure also shows the best-fitted Schechter-like function (dotted line) to the 
Bolshoi Planck / MultiDark Planck subhalo mass function of host halo mass 
$M_{\rm vir} = 10^{15}$\hMsun\ \citep[see][]{2016MNRAS.462..893R}. We highlight the remarkable 
agreement of the slope at the low-mass end, equal to -0.75, in agreement with previously 
reported works. Yet, the results obtained from Uchuu reveal a clear difference in the 
shape at the exponential tail, thanks to our superb statistics, with a shallow decline 
towards higher $M_{\rm acc}/M_{\rm vir}$ values. The $\sim20$ per cent difference 
in normalisation is due to the difference in the adopted cosmological parameters 
among both simulation suites.
A detailed study and interpretation of these results 
will be presented in a forthcoming paper (Molin\'e et al., in preparation).

\subsection{Mass-concentration relation: all halos}\label{sec:m-c}

The mass-concentration relation is an essential diagnostic to characterise
the internal structure of halos. Many past studies have investigated the dependence of 
concentration on halo mass and redshift
\citep[e.g.,][]{Prada2012, Ludlow2012, Dutton2014, Correa2015b, Klypin2016, Okoli2016,
  Pilipenko2017, Child2018, Diemer2019, Ishiyama2020}. 
These works show that halo concentration increases with decreasing halo mass and redshift, 
with the exception of the most massive and rare halos for which the concentration may 
increase with increasing mass \citep[e.g.,][]{Prada2012, Klypin2016,Diemer2019}.
Halo concentration also depends on the background cosmology and the relaxation state of halos 
\citep[e.g.,][]{Dutton2014, Klypin2016}, with relaxed halos exhibiting slightly larger 
concentrations.

In this paper, we extend these past works  
thanks to the statistical power of the Uchuu simulation suite, which 
covers an unprecedented range in halo mass at high resolution. 
For halos with an NFW profile \citep{Navarro1997}, concentration is defined as the 
ratio of the halo radius to the characteristic radius $r_{\rm s}$ of the NFW profile. 
In general, there are different ways to define the halo radius and mass. Here we use 
two methods, with halos defined either by the radius $r_{\rm 200}$ where
the average spherical overdensity is 200 times the critical density, or by 
the virial overdensity that slightly evolves with the redshift. We will 
label quantities related to each of the definitions by the indexes "200" or "vir", respectively.
Unless otherwise noted, our main results are show with $r_{\rm 200}$.
An alternative definition, using $r_{500}$, is often used for galaxy cluster modeling, 
in which the average spherical overdensity is 500 times the critical density.

\begin{figure*}
\centering 
\includegraphics[width=18cm]{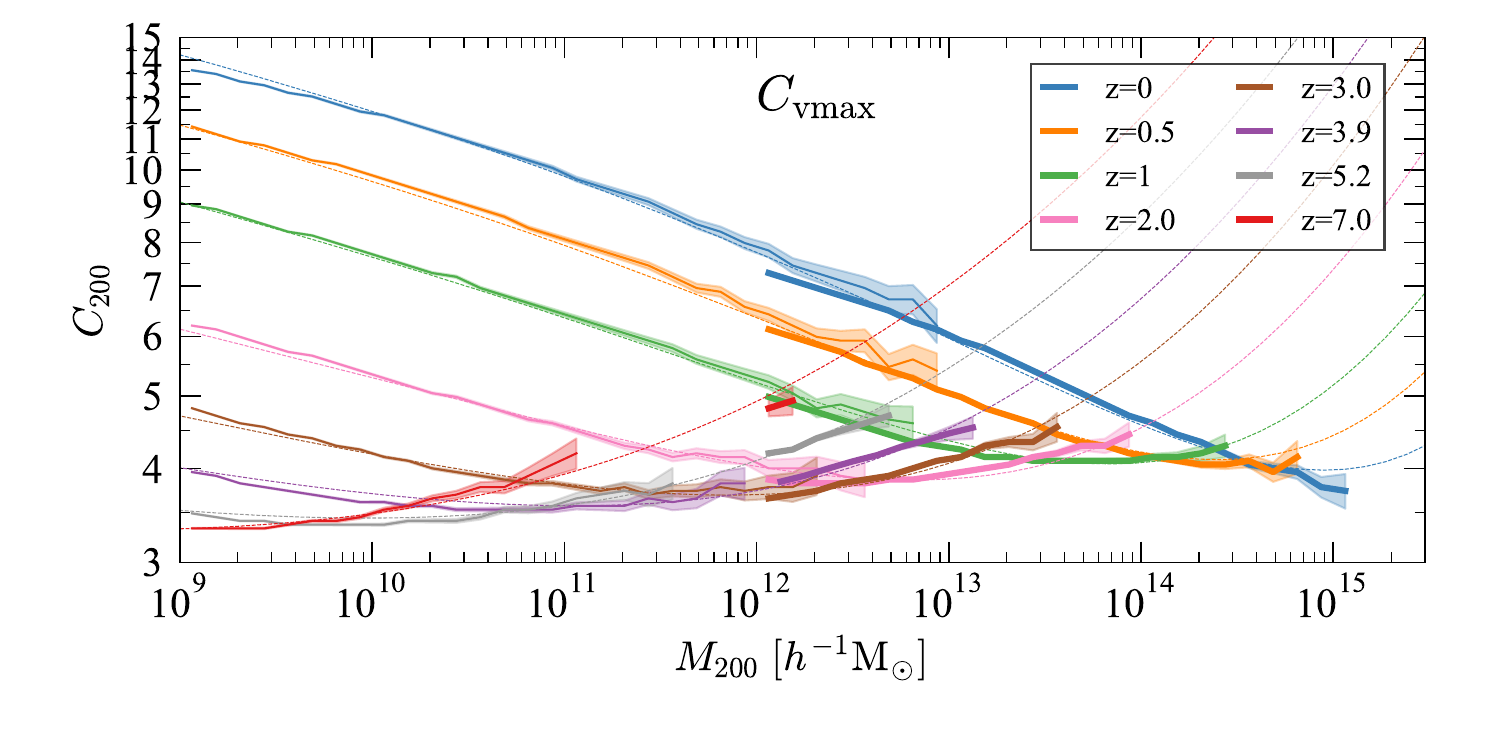} 
\includegraphics[width=18cm]{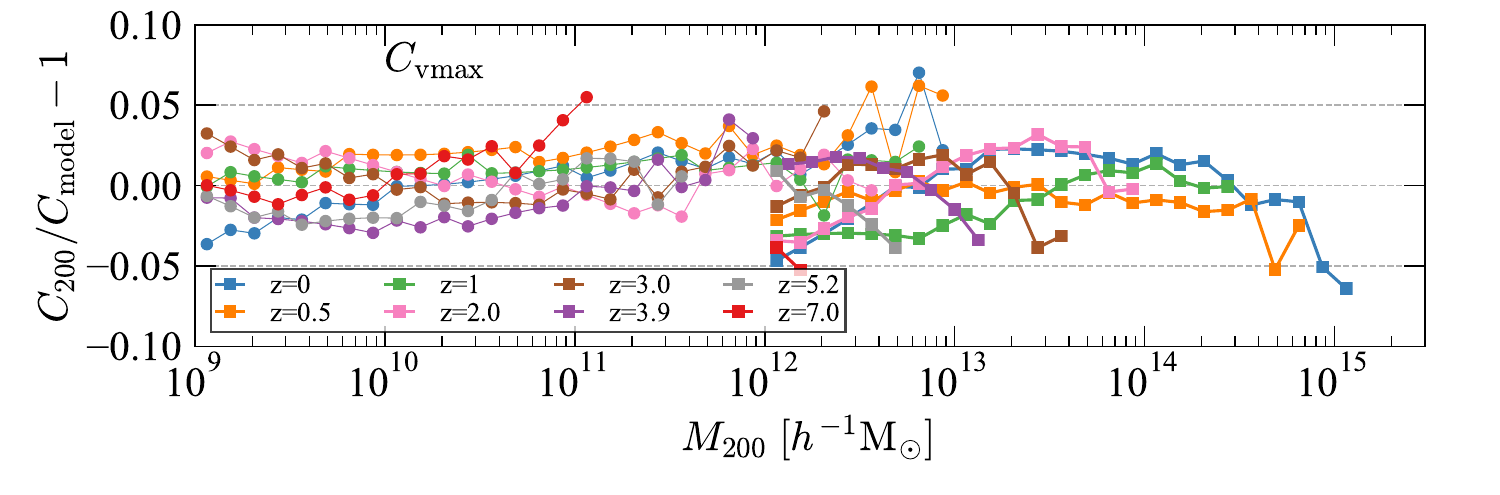} 
\caption{
{\it Top panel:}
Mass-concentration relation of halos 
for the Uchuu (thick curves) and the Shin-Uchuu (thin curves) simulations.
Median values with statistical uncertainties in each mass bin are shown. 
The concentrations are estimated using $V_{\rm max}$ given by eq.(\ref{eq:v200}).
The dashed curves are predictions of the analytical model described by eq.~(\ref{eq:Diemer}) 
in Appendix~\ref{sec:m-c_model}.
{\it Bottom panel:} 
The fractional difference between the halo concentrations of the Uchuu (squares) 
and the Shin-Uchuu (circles) simulations and the model predictions. It illustrates that 
the analytical model provides an accurate fit to the actual concentrations with an 
error less $\sim 5$ per cent, for the redshift range $0<z<7$ and halo masses covering 
six orders-of-magnitude.
}
\label{fig:m-c}
\end{figure*}

\begin{figure*}
\centering 
\includegraphics[width=18cm]{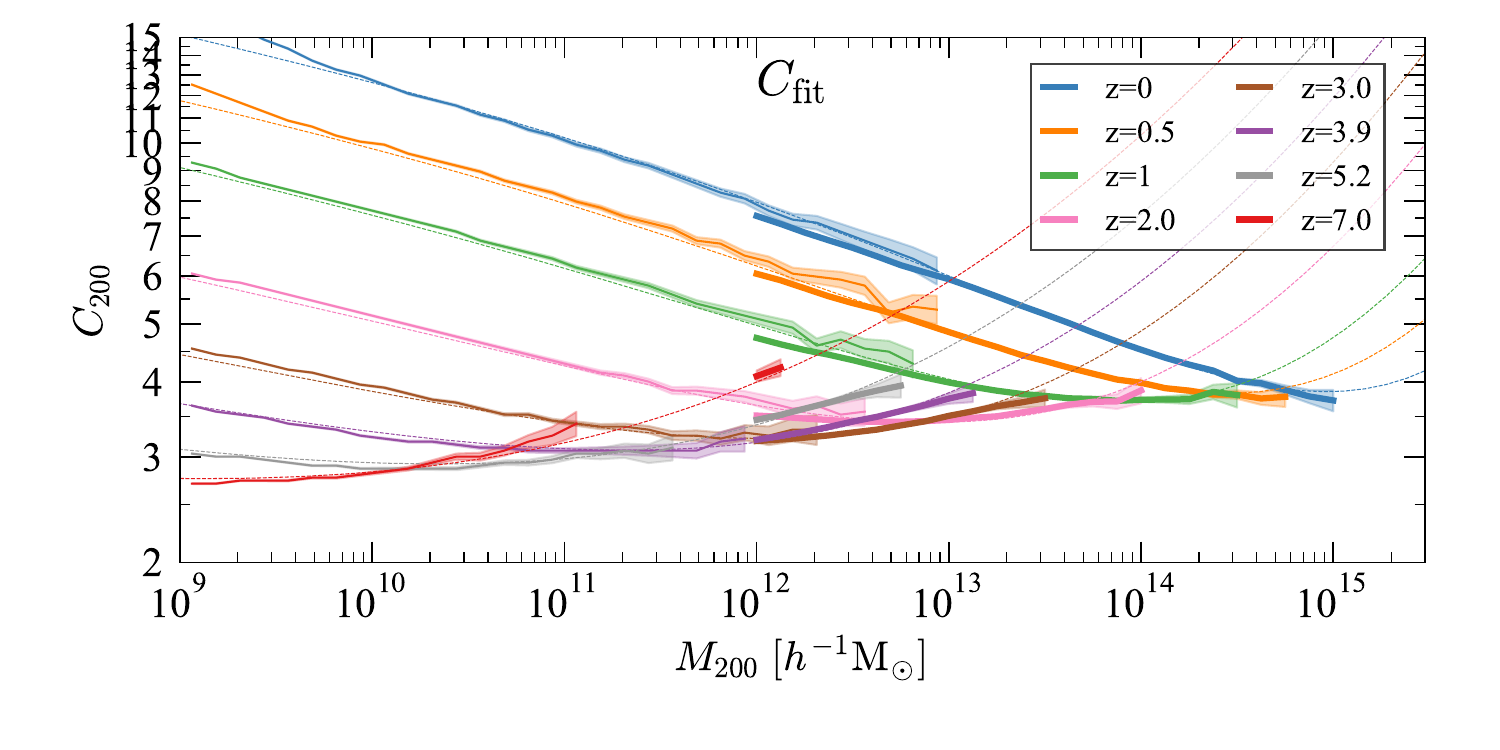} 
\includegraphics[width=18cm]{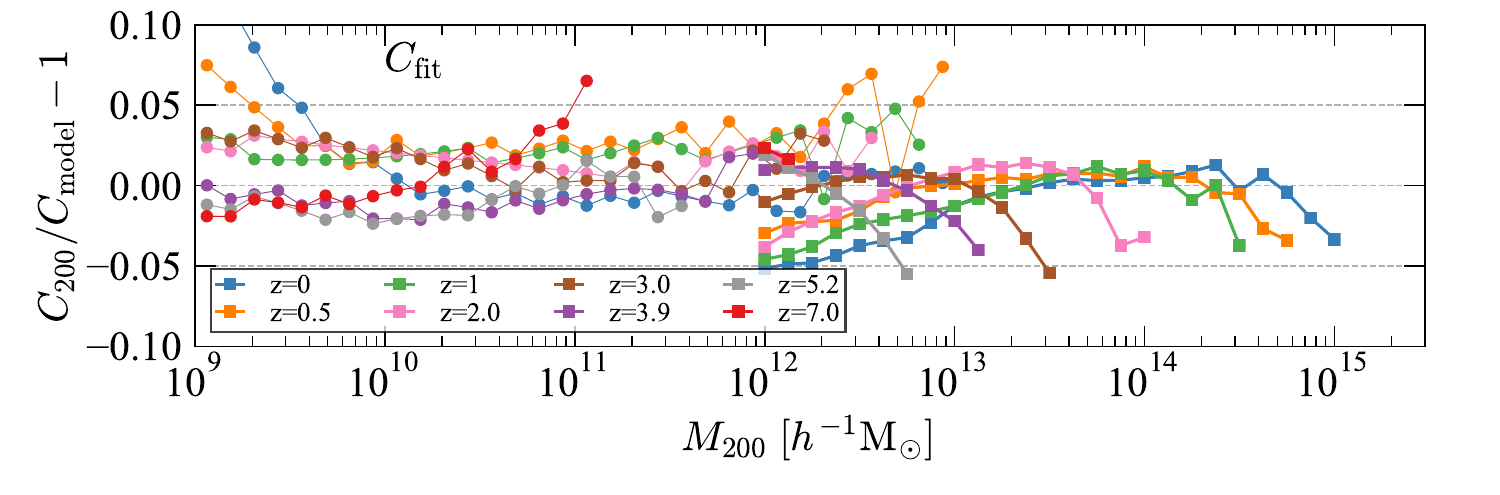} 
\caption{The same as in Figure~\ref{fig:m-c} but for concentrations estimated by 
profile fitting. The plot shows that the main tendencies of the concentration-mass
relation do not depend on the method used to find the concentration. This includes 
the upturn in concentration found at large masses.
}
\label{fig:m-c_rs}
\end{figure*}

\begin{figure}
\centering 
\includegraphics[width=9cm]{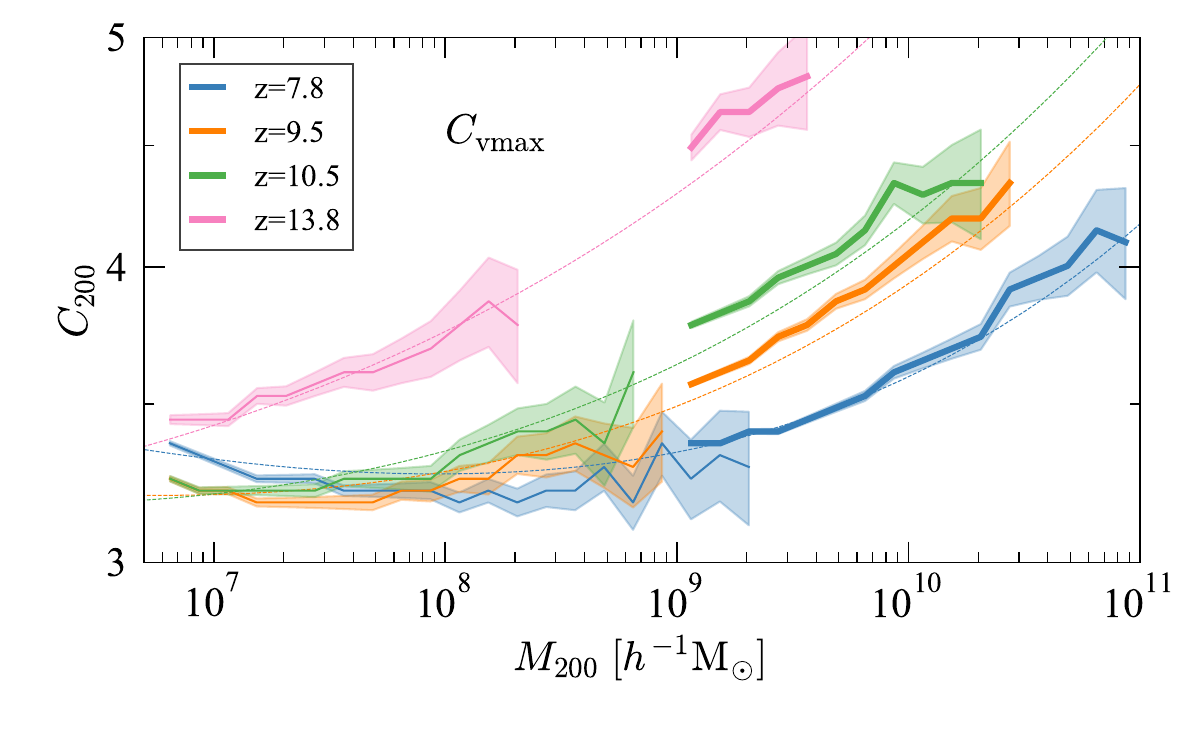}
\includegraphics[width=9cm]{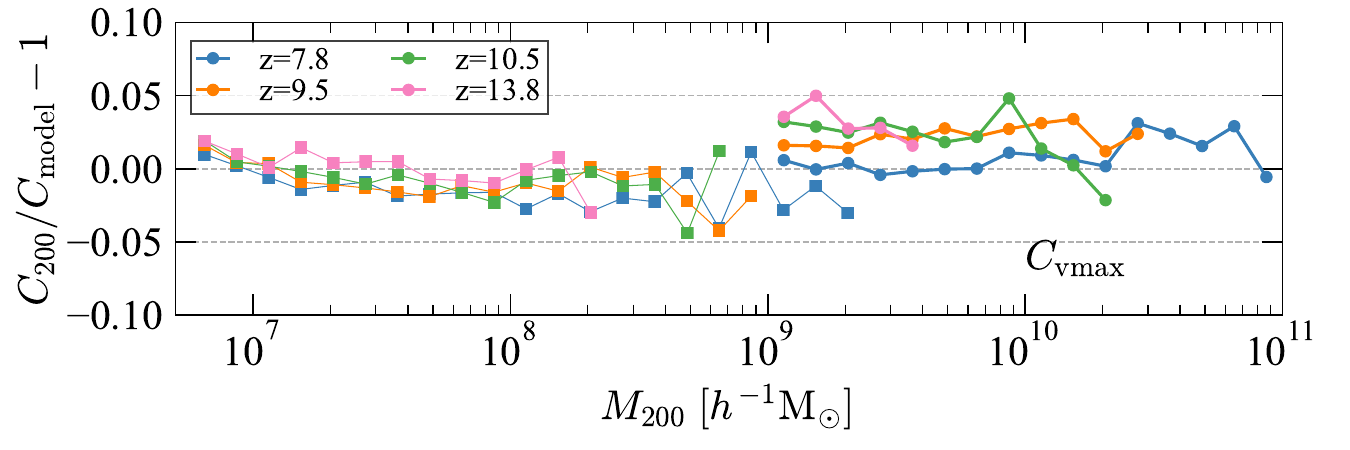}
\caption{
Halo concentrations at very large redshifts and small masses.
{\it Top panel:}
Mass-concentration relation for the Shin-Uchuu (thick curves) and the 
Phi-4096 (thin curves) simulations. In both cases, concentrations are estimated 
using the $V_{\rm max}$ method. Median values with statistical uncertainty in each 
mass bin are shown. The dashed curves provide the model prediction described by 
eq.~(\ref{eq:Diemer}). 
{\it Bottom panel:} 
The fractional difference between the halo concentrations of the Shin-Uchuu (circles) 
and the Phi-4096 (squares) simulations and the analytical model. Again, we find the 
model fits the simulation with an accuracy of 
better than $\sim 5$ per cent across a large redshift and mass range.
}
\label{fig:m-c_highz}
\end{figure}

\begin{figure}
\centering 
\includegraphics[width=9cm]{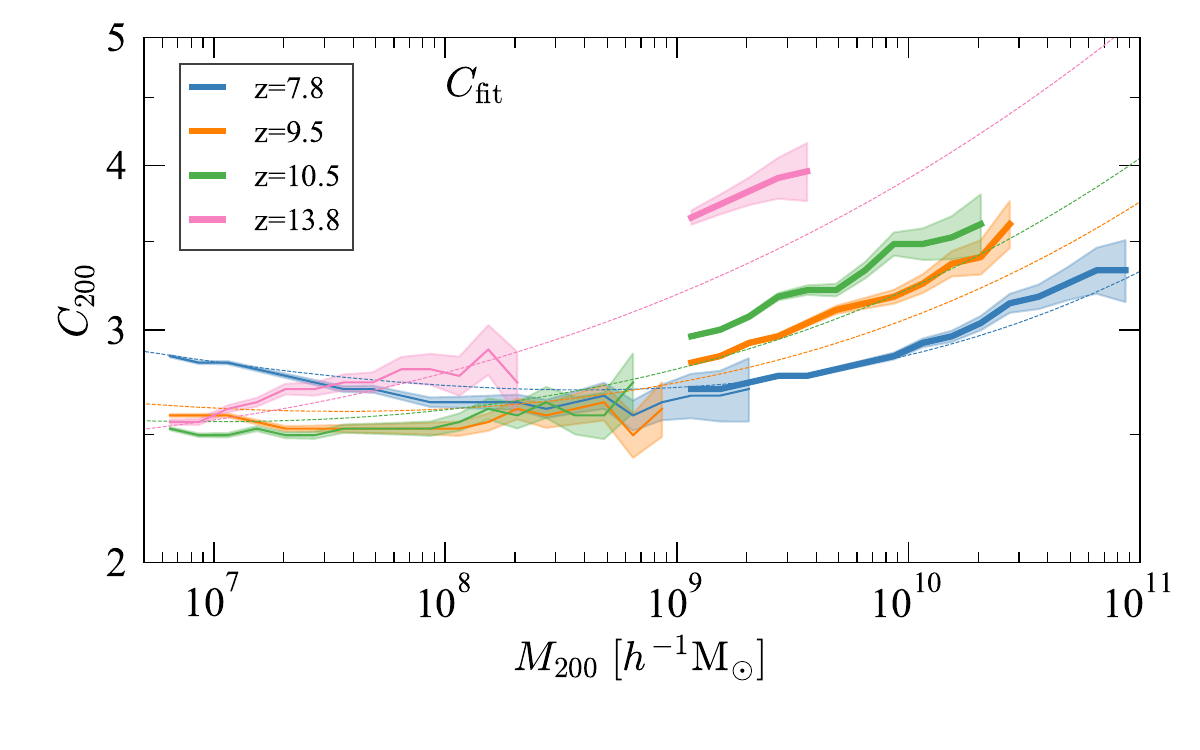}
\includegraphics[width=9cm]{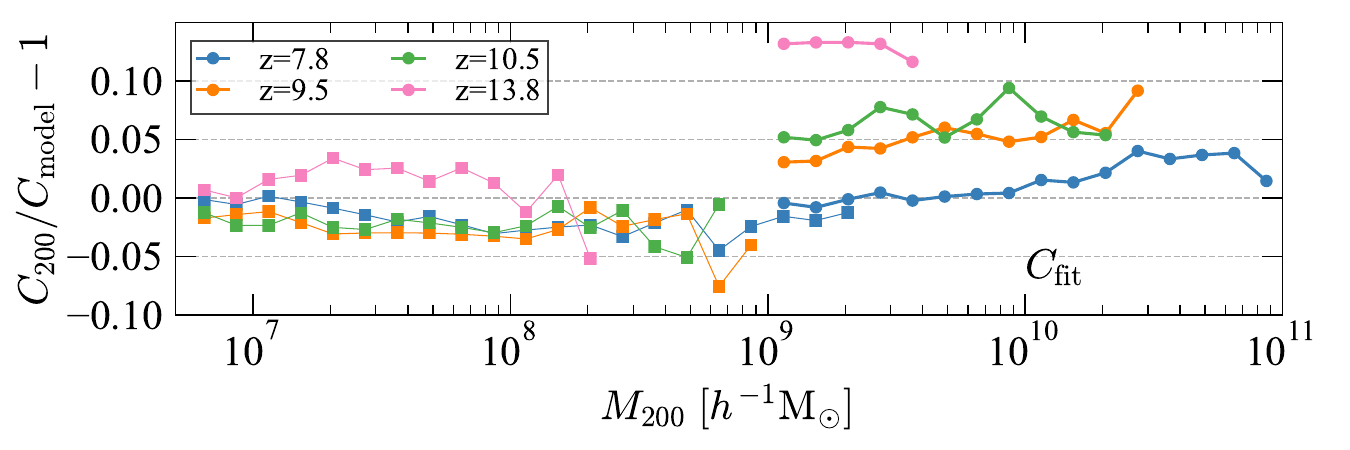}
\caption{
The same as in Figure~\ref{fig:m-c_highz}, but for concentrations estimated by profile 
fitting.
}
\label{fig:m-c_highz_rs}
\end{figure}

For halos well fit by an NFW profile, two parameters -- mass and concentration -- uniquely 
define the density distribution. The situation is more complicated for halos 
that do not follow an NFW profile. Indeed, some halos (and especially very massive and 
rare ones) are better described by the Einasto profile \citep{Navarro2004,Dutton2014,Klypin2016}. 
In this case one needs three parameters. If we still choose to use only two  
-- mass and concentration -- how do we then define the concentration? One must be careful when  
taking the naive association of the core radius as the radius $r_{-2}$ where the log-log 
slope of the density profile is -2, which is valid for the NFW profile. In this case two 
halos with obviously different concentrations may have the same $r_{-2}$ because 
they have different shape parameters of the Einasto profile. Indeed, \citet{Klypin2016} 
present examples that this typically happens with massive halos.
We similarly show cases of halo profiles of this kind in Appendix~\ref{sec:m-c_accuracy}. 
Keeping the above caveats in mind, for consistency in our analysis 
we use the same definition of concentration and the same algorithm as in 
the case of the two-parameter NFW profile. For such halos the accuracy 
of the prediction of their density distribution is reduced. However, in practice the 
errors for such halos are relatively small, even in extreme cases. For example, the 
density profile errors are less than 5 per cent for the most massive halos at $z=0$ at  
radii larger than 0.01 of the halo radius.

Halo concentration also depends on the dynamical state of the halo, 
with relaxed halos tending to have larger concentrations 
\citep[e.g.,][]{Prada2012, Ludlow2012, Klypin2016}. The selection criteria for 
relaxed halos can differ in literature, and can introduce systematic biases. 
For example, \citet{Prada2012} found an upturn in the concentrations of massive halos 
in both their "all" and "relaxed" samples, while \citet{Ludlow2012} did not find 
such an upturn in their relaxed ones. In this section we use all halos regardless of their relaxation state.
Relaxed halos and the upturn are discussed 
in the next section.

There are different methods that can be applied to the particle data to estimate halo 
concentration. One can construct and then fit the density profile in spherical 
shells. In this case the results can be affected by the quality of the fit and by the chosen
binning algorithm. Alternatively, we can define the concentration
by numerically solving the following equations \citep{Klypin2011, Prada2012}:
\begin{equation}
V_{200} = \left( \frac{GM_{200}}{R_{200}} \right)^{1/2}, \quad \frac{V_{\rm max}}{V_{200}} = \left( \frac{0.216c}{f(c)} \right)^{1/2}\label{eq:v200}, 
\end{equation}
where $f(c)=\ln(1+c)-c/(1+c)$ and $V_{\rm max}$ is the maximum circular velocity.
In this paper we use different notations to indicate these different methods. 
Concentrations obtained by fitting density profiles are labeled as 
$c_{\rm fit}$ while the concentrations from the $V_{\rm max}$ method are denoted as
$c_{\rm vmax}$. If no explicit label is given, the concentration was obtained with 
the $V_{\rm max}$ method.

Appendix~\ref{sec:m-c_accuracy} presents estimates of the errors for different methods and 
the dependencies of the estimates on numerical effects such as the number of particles and 
force resolution. There is a general trend that follows from the nature of the methods. 
The $V_{\rm max}$ is sensitive to the mass distribution in the relatively peripheral 
region of a halo and it is not sensitive to the mass distribution at the center 
or to any local fluctuations. This is a good method, if one is interested in 
the main body of the halo, but not its center. The density fitting method is 
more sensitive to the central regions. However, it is also prone to errors produced by 
non-relaxed features.

Both definitions produce similar values at lower redshifts, $z<1$, where halos tend to be 
more relaxed. On average, the differences in concentration are less than few per cent. 
However, when the fraction of non-relaxed halos increases (either at high redshifts or at 
the high-mass end) $c_{\rm fit}$ results in smaller concentrations \citep{Dutton2014}.  
At $z=5$, the difference between the two estimates can reach up to 15 per cent.
See Appendix~\ref{sec:m-c_accuracy} for more details.

When fitting our results with different analytical models we find that the model proposed by
\citet{Diemer2019} (with modifications) is the most accurate. However, their original 
calibration gave large errors that were $\sim 10$ per cent at low redshifts and $\sim 20$ 
per cent at high redshifts for concentrations from the $V_{\rm max}$ method. 
After tuning the parameters using Uchuu data covering 
large range of redshifts and masses, the final model showed significant improvements, 
even for concentrations based on $V_{\rm max}$. Following \citet{Diemer2019}, we assume that 
the halo concentration has the following functional form:
\begin{equation}
    c =  C\left(\alpha_{\mathrm{eff}}\right) \times \tilde{G}\left(\frac{A\left(\alpha_{\mathrm{eff}}\right)}{\nu}\left[1+\frac{\nu^{2}}{B\left(\alpha_{\mathrm{eff}}\right)}\right]\right),
    \label{eq:Diemer}
\end{equation}
where $\nu = \delta_{\rm c}/\sigma(M)$ is the height of the density peak, 
$\delta_{\rm c}=1.686$ is the critical overdensity for spherical collapse,  
and $\sigma(M)$ is the rms density fluctuation. Factor $n_{\rm eff}$ is the effective 
slope of the $\sigma(M)$ function, and $\alpha_{\rm eff}$ is the linear growth rate 
of fluctuations. Parameters $A$, $C$ and $B$ depend on $n_{\rm eff}$. The function $\tilde{G}$ 
depends on the assumed density profile (here NFW).
Appendix~\ref{sec:m-c_model} gives details of the approximation. 
Equation~\ref{eq:Diemer} has six free parameters, which we find by 
minimizing the errors using our Uchuu suite of simulations. 
Table~\ref{tab:fit_param} presents our parameters for 
the analytical model. We also present the parameters for concentrations 
$c_{\rm vir}$ estimated using the $V_{\rm max}$ method and profile fitting.

Figure~\ref{fig:m-c} (top panel) shows the mass-concentration relation of the
Uchuu and Shin-Uchuu simulations and their redshift evolution.
Median values with statistical uncertainty in each mass bin are shown.
The concentrations are estimated using the $V_{\rm max}$ method (eq.(\ref{eq:v200})).
At low redshift, the relation follows a well-known power law behaviour
except at the highest masses. We observe a flattening and an upturn with
increasing mass, consistent with previous findings
\citep[e.g.,][]{Prada2012}. This flattening and upturn gradually
shifts to lower mass at higher redshift.
The Shin-Uchuu simulation produces slightly larger concentrations than Uchuu 
in the overlap region ($10^{12}\sim10^{13}$\hMsun) at $z<3$. 
However, both simulations follow the same trend.
Figure~\ref{fig:m-c_rs} (top panel) also shows the mass-concentration relation, 
but where concentrations are estimated by the profile fitting method. 
Together, Figures~\ref{fig:m-c} and \ref{fig:m-c_rs} highlight that the main tendencies 
of the concentration-mass relation does not depend on the method to find concentration
in our simulation suite. This includes the upturn in the concentration at large masses.

With the unprecedented high resolution and
statistical power of the Shin-Uchuu and the Phi-4096 simulations, we
can statistically study the mass-concentration relation at 
mass $M_{200} < 10^{11}$\hMsun\ and redshift $z>7$ for the first time.  
The top panels of Figures~\ref{fig:m-c_highz} and \ref{fig:m-c_highz_rs} 
show the mass-concentration relation of the
Shin-Uchuu and Phi-4096 simulations and their redshift evolution, 
using the $V_{\rm max}$ and profile fitting methods, respectively.
For both, concentration increases with increasing mass, 
opposite to the behavior see at lower redshift.
This should be compared to the flattening and upturn 
identified in Figures~\ref{fig:m-c} and \ref{fig:m-c_rs}.
Again, the main trends of the concentration-mass
relations do not depend on the method to find concentration.

The bottom panel of both Figures~\ref{fig:m-c} and \ref{fig:m-c_highz} show 
residuals of the concentration estimated using the $V_{\rm max}$ method relative 
to the model. Across most of the halo mass and
redshift range the difference is within 5 per cent, although there are a few exceptions.
This highlights that the accuracy provided by the model is quite good 
for masses spanning nearly eight orders of magnitude. 
Similarly, the bottom panel of both Figures~\ref{fig:m-c_rs}
and \ref{fig:m-c_highz_rs} show residuals using the profile fitting method.  
The model accuracy here is also good but slightly worse than 
for concentrations estimated using $V_{\rm max}$. 
Again, the difference is within 5 per cent for most halo masses and redshifts,
although it can increase to nearly 10 per cent in places.
This is prominent for rare objects at high redshift. 

We also explore the concentration relation using $r_{500}$: $\rm c_{500} = r_{500}/r_s$.
The same fitting procedure described in this Section is used 
to find the best fitting parameters of the mass-concentration relation.
Plots for concentration-mass relation are presented in Figure~\ref{fig:m-c_relaxed}. 
We present concentration estimated using only the profile fitting 
because it is better to describe the halo inner profile 
than the $V_{\rm max}$ method as shown in Appendix~\ref{sec:m-c_accuracy}.
We use only data at $z \le 7$ to ensure the fitting error within 5 per cent. 
Including data at $z>7$, the fitting error increases from 5 to 10 per cent.

Table~\ref{tab:fit_param} presents the parameters for the $c_{200}$, $c_{\rm vir}$ and $c_{500}$
halo definitions estimated using the $V_{\rm max}$ and profile fitting methods.
In general, the model accuracy is very similar for $c_{200}$;  
the difference is within 5 per cent for most halo masses and redshifts, 
but is close to 10 per cent for rare objects at high redshift
when concentrations are estimated by profile fitting. 

\begin{table}
\centering
\caption
{
Best fitting parameters of the mass-concentration relation 
for the model eq.~(\ref{eq:Diemer}). Parameters are given for
different methods of finding concentrations, for different halo selections, and for
different halo definitions.
We present concentration $c_{500}$ estimated using only the profile fitting 
because it is better to describe the halo inner profile 
than the $V_{\rm max}$ method as shown in Appendix~\ref{sec:m-c_accuracy}.
}
\label{tab:fit_param}
\begin{tabular}{lcccccc}
\hline
 $c$ & $\kappa$ & $a_0$ & $a_1$ & $b_0$ & $b_1$ & $c_\alpha$\\
\hline
\multicolumn{7}{c}{Vmax\,\, All halos}\\
 $c_{200}$ & 1.10 & 2.30 & 1.64 & 1.72 & 3.60 & 0.32 \\
 $c_{\rm vir}$ & 0.76 & 2.34 & 1.82 & 1.83 & 3.52 & $-0.18$ \\
\hline
\multicolumn{7}{c}{Fit\,\, All halos}\\
$c_{200}$ & 1.19 & 2.54 & 1.33 & 4.04 & 1.21 & 0.22 \\
$c_{\rm vir}$ & 1.64 & 2.67 & 1.23 & 3.92 & 1.30 & $-0.19$ \\
$c_{500}$   & 1.83 & 1.95 & 1.17 & 3.57 & 0.91 & 0.26 \\
\hline 
\multicolumn{7}{c}{Vmax\,\, Relaxed halos}\\
$c_{200}$ & 1.79 & 2.15 & 2.06 & 0.88 & 9.24 & 0.51 \\
$c_{\rm vir}$ & 2.40 & 2.27 & 1.80 & 0.56 & 13.24 & 0.079 \\
\hline
\multicolumn{7}{c}{Fit\,\, Relaxed halos}\\
 $c_{200}$ & 0.60 & 2.14 & 2.63 & 1.69 & 6.36 & 0.37 \\
 $c_{\rm vir}$ & 1.22 & 2.52 & 1.87 & 2.13 & 4.19 & $-0.017$ \\
  $c_{500}$  & 0.38 & 1.44 & 3.41 & 2.86 & 2.99 & 0.42 \\
\hline
\end{tabular}
\end{table}

\begin{figure}
\centering 
\includegraphics[width=9cm]{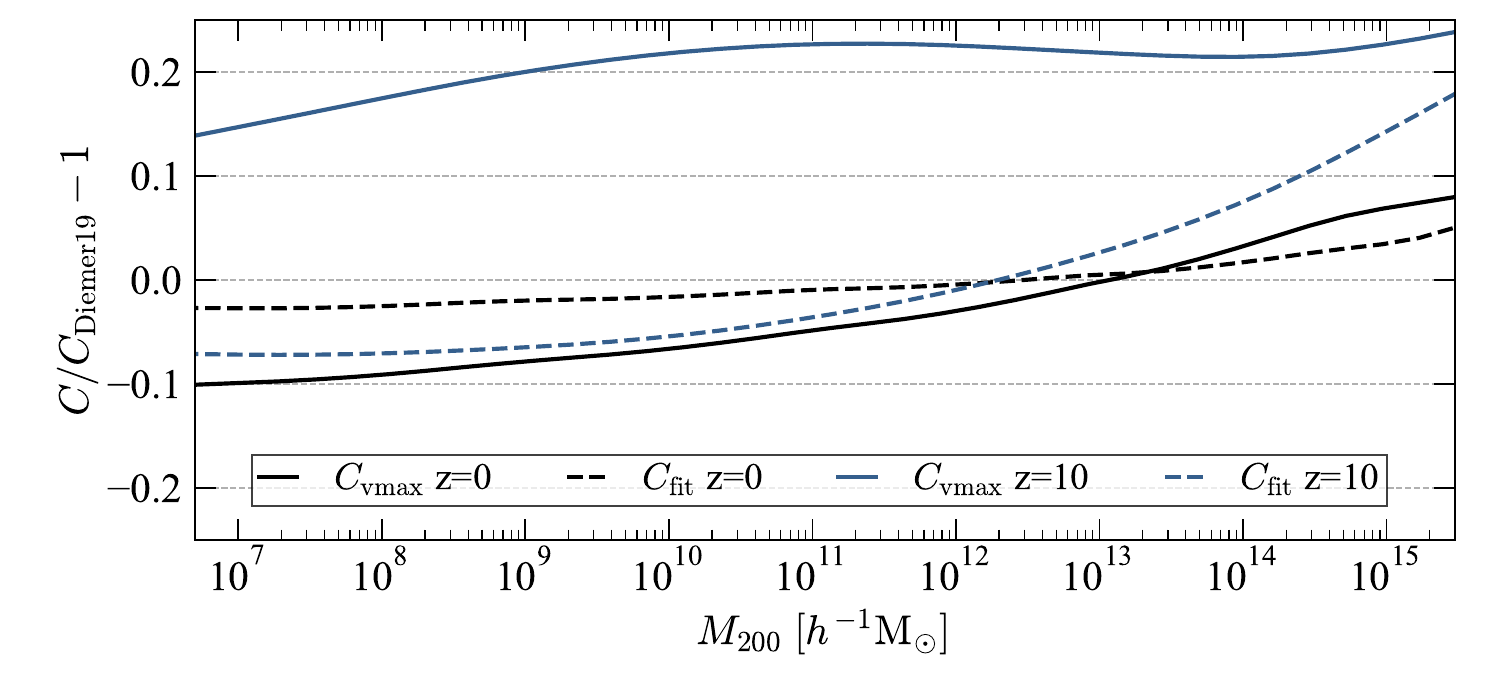} 
\caption{
Comparison of the predictions of halo concentrations $c_{200}$ in this paper with those of \citet{Diemer2019}. 
Both predictions use the same functional form, but different parameters.
}
\label{fig:Diemer}
\end{figure}

In spite of the fact that we use the same functional form proposed by \citet{Diemer2019}, 
our parameters of the approximation are quite different from theirs.
Figure~\ref{fig:Diemer} compares predictions for such concentrations. The agreement between the 
predictions is remarkably good for $c_{\rm fit}$ estimates at $z=0$ with just $\sim 2$ per cent 
differences for masses $M<5\times 10^{14}$\hMsun. The differences increase at high
redshifts and become $\sim 20$ per cent at $z=10$. This is a good improvement as compared with 
the situation just a few years ago. For example, disagreement between \citet{Correa2015b} and 
\citet{Diemer2015} was 20 per cent at $z=0$, $M=2\times 10^{14}$\hMsun. A more detailed 
comparison of results provided by different publications can be found in \citet{Diemer2019}.

\subsection{Mass-concentration relation: relaxed halos}\label{sec:m-crel}

So far we have presented mass-concentration relations for all halos regardless of their dynamical state. 
For some studies one may need to remove rare transient events (e.g., due to ongoing major mergers) and 
to study more relaxed halos \citep[e.g.,][]{Neto2007,Prada2012, Ludlow2012, Klypin2016}. 
There are no clear selection criteria of relaxed halos. 

The first obvious choice is the virial parameter $2K/|W|-1$, where $W$ and $K$ are
the potential and kinetic energies, respectively. The virial parameter should be equal to zero 
for objects in virial equilibrium. As such, this parameter has been extensively used in the past
\citep[e.g.,][]{Neto2007,Ludlow2012}. However, many results show that otherwise relaxed halos can have 
systematically too large a $2K/|W|-1$, as if the kinetic energy is too big for a given potential energy. 
The problem is extreme with values $2K/|W|-1>0.3$ being typical for halos at redshifts $z>3$ 
\citep{Davis2011,Klypin2016}. If this were true, halos would have dissolved or collapsed on a 
dynamical time-scale. Analysis of individual and average halo profiles do not show any signs of 
catastrophic collapse or expansion: inner halo regions have zero average radial velocities while 
there are small infalling velocities close to the virial radius. This failure of the virial 
parameter is attributed to simplistic treatment of the virial theorem. Dark matter halos are not 
isolated objects. When applying the virial theorem, one must include the surface pressure term, which, 
for spherical halos, is $S_{\rm p}=4\pi R^3_{\rm vir}\rho_{\rm vir}v_{\rm r}^2$, where $v_{\rm r}^2$ is 
the dispersion of the radial velocity. Once this correction is applied, the virial parameter becomes 
significantly smaller \citep{Shaw2006,Davis2011,Klypin2016}. Furthermore, large and complicated corrections 
and uncertainties due to non-spherical halo shapes make the virial parameter much less useful as an 
indicator of non-equilibrium halos. This is why the virial parameter is currently either not utilized, 
or used with a very large upper limit \citep{Correa2015b,Klypin2016,Child2018}.

There are other conditions for selecting relaxed halos. These include the fraction of mass in 
subhalos $f_{\rm sub}$, the offset parameter $X_{\rm off}$, and the spin parameter $\lambda$.
We define the offset parameter as the distance between the center of halos and center of mass: 
$X_{\rm off}=|\Vec{r}_{\rm center}-\Vec{r}_{\rm cm}|/R_{\rm vir}$. The spin parameter is defined 
as $\lambda = J|E|^{1/2}/GM^{5/2}_{\rm vir}$, where $E$ is the total energy and $J$ is the angular momentum. 

None of these parameters actually measure deviations from equilibrium, mostly because it is difficult to
precisely specify what a deviation from equilibrium is. Currently used parameters in the literature all are 
related to the amount of substructure. For example, an on-going major merger may result in a large 
offset parameter. As the merger proceeds, the subhalo is tidally stripped and finally merges with the 
parent halo resulting in the decline of $X_{\rm off}$. A combination of restrictions on $X_{\rm off}$ 
and $\lambda$ typically removes large merging events. In this paper we accept a set of conditions advocated by
\citet{Klypin2016}, who have found that the combination of the following three conditions
are a reasonable choice to select relaxed halos:
\begin{equation}
2K/|W| < 1.5, \quad X_{\rm off} < 0.07, \quad \rm and \quad \lambda < 0.07. \label{eq:relaxation}
\end{equation}
Applying the same fitting procedure as described in \S~\ref{sec:m-c}, we provide best
fitting parameters of the mass-concentration relation of relaxed halos in Table~\ref{tab:fit_param}.

\begin{figure}
\centering 
\includegraphics[width=7.8cm]{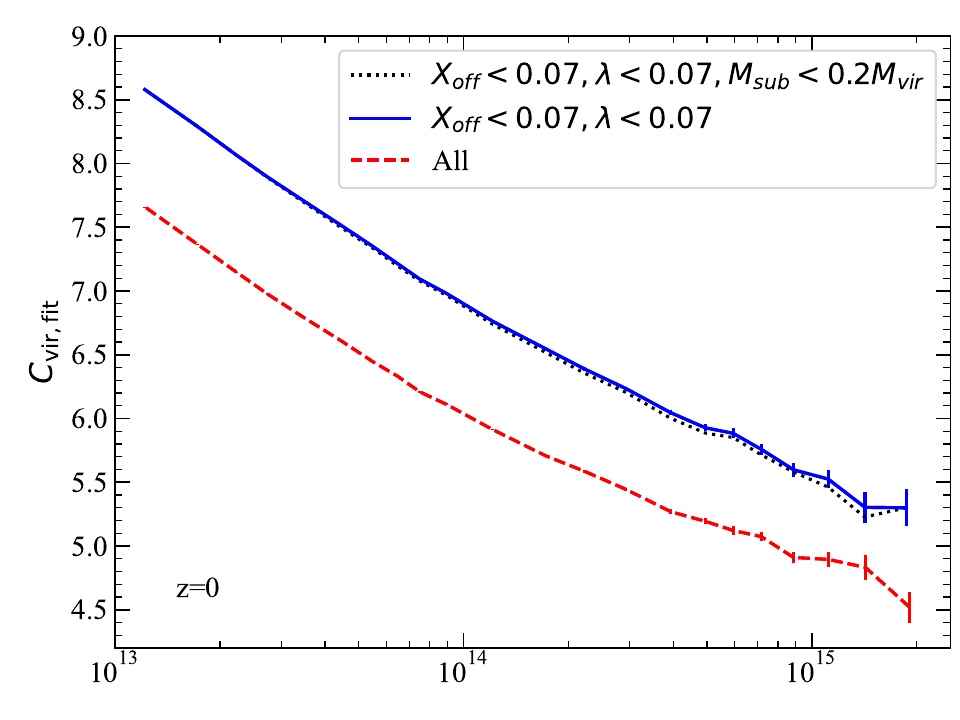} 
\includegraphics[width=7.8cm]{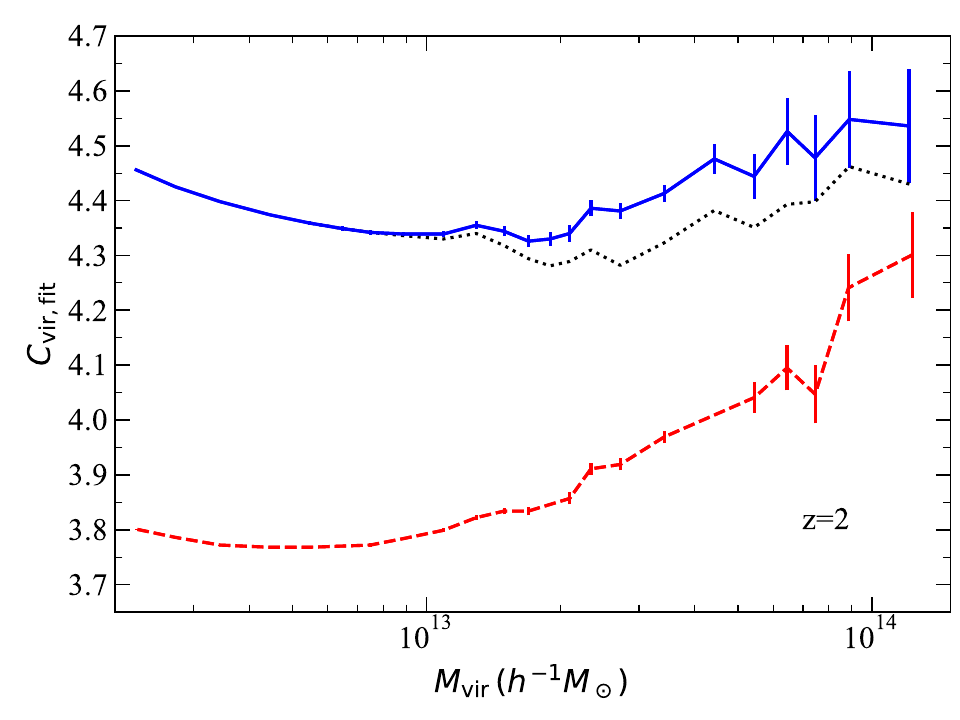} 
\caption{
Comparison of concentrations of massive relaxed and all halos at $z=0$ (top panel) and $z=2$ (bottom panel). 
Here we use virial masses and concentrations. The full curves show concentrations of halos restricted by the 
offset parameter $X_{\rm off}< 0.07$ and spin parameter $\lambda < 0.07$.  The dotted curves are
for halos with an extra condition: we remove halos with the most massive subhalo exceeding 20 per cent of 
the mass of the central halo. The error bars show the statistical uncertainties of the average values of 
the concentrations. Note the dramatic difference between $z=0$ and $z=2$ concentrations: no upturn at $z=0$ 
and a prominent upturn at $z=2$. The upturn is weaker but still present for relaxed halos regardless of 
which of the two sets of conditions are used. For all redshifts and halo masses, halo concentration is 
{\it larger} for relaxed halos compared with the full population.
}
\label{fig:relaxedC-M}
\end{figure}

Figure~\ref{fig:relaxedC-M} compares concentrations of relaxed and all halos at $z=0$ and $z=2$ (solid curves). 
Here we focus on the high-mass part of the concentration-mass relation. Appendix~\ref{sec:m-c_relaxed} presents 
further plots of the concentration of relaxed halos. We also use another set of conditions to select 
relaxed halos: in addition to the main conditions in Equation~(\ref{eq:relaxation}) we also remove halos 
if the mass of the largest subhalo exceeds 20 per cent of halo mass (dotted curves). 

The most notable feature in this figure is the dramatic difference between $z=0$ and $z=2$ concentrations. 
At $z=0$ there is no indication of an upturn in the concentration-mass relation, though the relation is not a power-law 
as is the case for much smaller halos. At $z=2$ there is a prominent upturn in the population of all halos. 
The upturn is weaker but still present for relaxed halos regardless of which of the two sets of conditions 
are used. 

For all redshifts and for all halo masses halo concentration is {\it larger} for relaxed halos. The 
additional constraint on the mass of the largest subhalo $f_{\rm sub}<0.2$ does not affect much 
the concentration-mass relation. It is barely noticeable at $z=0$. At $z=2$ the effect is larger but 
still small. There is a good reason for this: conditions for the offset $X_{\rm off}<0.07$ and spin 
$\lambda<0.07$ have already removed most of the major-merger events. This can be clearly seen from 
the fraction of relaxed halos. For example, at $z=2$ the fraction of relaxed halos with masses 
$M_{\rm vir}>8\times 10^{13}$\hMsun\ is 43 per cent of the total 900 halos. Adding the constraint on 
subhalos only removes another 5 per cent of halos. We also tried to mimic one of the selection conditions 
used by \citet{Ludlow2012} by removing halos having their total mass of subhalos larger than 10 per cent 
of the halo mass. This reduces the fraction of relaxed halos to 30 per cent without much of an effect 
on the upturn (we find only an additional 1.7 per cent decline of halo concentration at the upturn).

One of the contentious issues in this field is the existence and nature of an upturn 
in the halo concentration--halo mass relation at the massive end. It was first discovered 
by \citet{Klypin2011} and later confirmed by \citet{Prada2012,Dutton2014,Diemer2015,Klypin2016,Diemer2019}. 
There is a general consensus that the phenomenon is real as far as the population of all halos 
is concerned. However, there are widely different opinions regarding its nature.

Analytical models for the evolution of halo concentrations so far do not help us understand the nature 
of the upturn. Those used by \citet{Prada2012} and \citet{Diemer2015,Diemer2019} assume the upturn,  
so it is not an argument in favor of it. Models that relate the evolution of the halo concentration 
with the mass accretion history \citep{Bullock2001,Zhao2009,Correa2015b} {\it assume} that the concentration 
increases as the halo accretes mass after an episode of major merger. Again, these models do not predict the upturn.
The problem is that mass accretion history is not the only factor that defines halo concentration. 
Simulations of \citet{Klypin2016} clearly show that halos at the upturn have much more radial orbits 
when compared with halos at the declining branch of the concentration-mass relation (see, for example, Figure 14 
in \citet{Klypin2016}). In turn, large radial velocities of dark matter particles result in more mass reaching
small radii, which is the reason for increased concentration. We find the same effect in the Uchuu simulation. 

Finding the upturn in cosmological simulations is a difficult numerical problem because it requires relatively 
high mass resolution and a very large simulation box to have enough statistics of massive halos. For example,
\citet{Correa2015b} did not find the upturn in either the total population or in relaxed halos.
However, their largest computational box was only 400\hMpc. The only chance for them to see the upturn was 
at $z=2$: at lower redshifts the upturn is weaker and at higher redshifts their computational box and mass 
resolution are too small to resolve the required population. The largest halos at $z=2$ in \citet{Correa2015b} 
had mass $\sim 10^{13}$\hMsun, and statistical uncertainties for the concentration of these 
halos was $\sim 10$ percent. They did not see the upturn because it starts at masses larger than $M=10^{13}$\hMsun\
(see Figures \ref{fig:m-c} and \ref{fig:relaxedC-M}). 

A similar situation happens for simulations of \citet{Dutton2014} who had larger box of 1\hGpc\ 
(with thus improved statistics), but a mass resolution of $4\times 10^{11}$\hMsun. Thus, again we do not expect
them to see the upturn because their largest halo of $M=2\times 10^{13}$\hMsun\ had only 50 particles -- 
not enough to estimate the concentration. Other boxes with better mass resolution were too small to 
definitively find the upturn. 

Our results for relaxed halos contradict those of \citet{Ludlow2012} who argue that once unrelaxed halos are 
removed, the concentration of the remaining massive halos {\it declines}, resulting in the disappearance of 
the upturn (see their Figure 1). When we remove unrelaxed halos in Uchuu simulation, the concentration 
{\it increases} for all halos including those in the upturn. It is interesting to explore why we 
have such disagreements. \citet{Ludlow2012} used three conditions to select "relaxed halos": 
$2K/|W| < 1.3$, $X_{\rm off} < 0.07$ and the fraction of mass in subhalos $f_{\rm sub}<0.1$. 
At first sight, these conditions are similar to what we use in Figure~\ref{fig:relaxedC-M}. 
As discussed above, as a test we also implemented the condition $f_{\rm sub}<0.1$, 
but it did not make much of a difference. In spite of the seemingly similar conditions the overall 
disagreement with \citet{Ludlow2012} is quite dramatic. This can be illustrated by comparing the fractions 
of relaxed halos at $z=2-3$. In Uchuu simulation, this fraction is about 30 per cent once we add a constraint 
on the mass of subhalos. In the case of \citet{Ludlow2012}, their fraction of relaxed halos was significantly 
smaller: 2 per cent at $z=3$ and 11 percent at $z=1$. The reason for the small fraction of relaxed halos 
is due to smaller cutoff of the virial parameter used by \citet{Ludlow2012}. 
As discussed in Sec.~\ref{sec:m-crel}, without a correction for the surface pressure the virial parameter 
provides very biased results, giving the impression that most of the high-redshift massive halos are out of 
virial equilibrium, while in reality they are not. Figures~5 and 6 in \citet{Klypin2016} show the effect of this 
surface pressure correction, which has a strong dependence on redshift. At $z=3$, the vast majority of 
halos would violate the condition $2K/|W| < 1.3$ used by \citet{Ludlow2012}, and thus are considered unrelaxed. 
After the correction, most of them pass this condition and become "relaxed enough" again. Thus we can conclude
that the disagreement with \citet{Ludlow2012} is due to an unrealistic constraint on the virial parameter 
used in their analysis.

We can also now settle another issue: is there an upturn at $z=0$? The argument for the presence of late-redshift 
upturn was always statistically questionable. With the large volume of the Uchuu simulation we can clearly state 
that there is no $z=0$ upturn. 

\section{Summary and Future Prospects}\label{sec:summary}

Ongoing and upcoming wide and deep surveys will provide vast amounts of data on galaxies 
and AGN over Gpc scales. To get as much information as possible from such surveys, 
precise mock galaxy and AGN catalogues are needed. For this purpose, we conducted 
a suite of ultra-large cosmological {\it N}-body simulations, the Uchuu simulation suite,  
and constructed halo/subhalo merger trees for them.
The largest simulation in the suite, Uchuu, consists of $12800^3$ dark matter particles
in a box of 2.0\hGpc, with particle mass $3.27 \times 10^{8}$\hMsun.  
The highest resolution simulation in the suite, Shin-Uchuu,  
consists of $6400^3$ particles in a 140\hMpc\ box, 
with particle mass $8.97 \times 10^{5}$\hMsun.
By combining these simulations, we can follow the evolution of dark matter halos 
(and subhalos) spanning the mass range from dwarf galaxies to massive galaxy clusters.
Thanks to its large volume and high mass resolution, our simulation suite provides 
one of the the most accurate theoretical templates to date to construct galaxy and 
AGN catalogues.

In this paper, we have presented the basic properties of the Uchuu suite, which includes
halo demographics, the matter power spectra, and halo and subhalo mass functions. 
These demonstrate the very large dynamic range and superb statistics of the Uchuu simulations.
Then we provided an accurate model of halo concentrations that describes our simulations.
The results are summarized as follows.

\begin{enumerate}
\item 
From the analysis of the evolution of power spectra, we have demonstrated that
our simulations are accurate for use across a wide dynamic range, from the Baryon Acoustic
Oscillation scale down to very small structures.
\item
We have shown the halo mass function with a mass range spanning 
approximately eight orders of magnitude, reflecting the power of our
very large simulation set.  The convergence between simulations is
remarkably good -- within a few per cent -- ensuring the accuracy of our
results.
\item
The subhalo mass function, with $M_{\rm acc}/M_{\rm vir}$ ranging about six orders 
of magnitude, is in good agreement with previous simulation work and various model 
predictions. Yet, our superb statistics reveal a clear difference  
at the high-mass exponential tail, with a shallow decline towards higher 
$M_{\rm acc}/M_{\rm vir}$ values.
\item
Using the Uchuu suite, we have calibrated the parameters of the halo concentration 
evolution model presented in \citet{Diemer2019}. Our fitting errors are within 5 
per cent for halos with masses spanning nearly eight orders of magnitude and 
covering the redshift range $0 \leq z \leq 14$.
\item 
We confirm that there is an upturn in the mass-concentration relation for the population 
of all halos and of relaxed halos at redshifts $z\gtrsim 0.5$.  There is no upturn
at lower redshifts.
\end{enumerate}

The data products used in this paper form part of the Uchuu public data release 1, 
and can be found at \urluchuu. This includes subsets of simulation particles, matter 
power spectra, halo/subhalo catalogues, and their merger trees. The latter will be described
in an upcoming publication, and are produced by postprocessing the data with the well  
established tools \textsc{rockstar} \citep{Behroozi2013} and \textsc{consistent trees} 
\citep{Behroozi2013b}. Particle data is provided in the \textsc{gadget-ii} format 
\citep{Springel2005b}; selected other products use the \textsc{hdf5} format to help reduce 
reading and analysis times. This public data release increases the accessibility and 
portability of our data provides new opportunities for scientific exploitation. 

The superb numerical resolution achieved by the Uchuu simulations over an extremely wide halo mass 
range makes them uniquely powerful to characterize the subhalo population in great detail as well. 
Indeed, in a forthcoming publication (Molin\'e et al. in preparation)
the full set of Uchuu simulations in Table~\ref{tab:sim}, together with the Phi-4096 high-resolution simulation described in Table~\ref{tab:sim}, 
will be used to study subhalo concentrations, abundances and distributions. Potential dependencies 
of subhalo properties with, e.g., host halo mass or distance to host halo centre will also be explored. 
All together, we anticipate that Uchuu and Phi-4096 will allow us to test, under the same consistent 
framework, subhalo maximum circular velocities as low as $\rm \sim1~km/s$ and as high as $\rm \sim1800~km/s$ 
or, equivalently, subhalo masses between $\sim10^{5}-10^{15}$~$h^{-1}M_\odot$, with unprecedented statistics. 
This impressive subhalo mass range will thus enable the study of halo hierarchy over several orders of 
magnitude, indeed reaching subhalos as light as $10^{-7}$ the mass of their hosts for some halos. 
In addition to its inherent cosmological value, this upcoming work on Uchuu halo substructure will be 
of particular interest to a large variety of topics currently under debate and scrutiny in adjacent 
fields. For instance, it may serve to elucidate the precise role of subhalos for dark matter searches,
e.g.,~\cite{2012ApJ...747..121A,Sanchez-Conde14,Moline2017,Coronado-Blazquez19a,Ishiyama2020}; 
or to shed further light on the actual survival of subhalos within their hosts 
\citep{2018MNRAS.474.3043V,2018MNRAS.475.4066V, Meneghetti2020}.

The Uchuu suite of simulations will be used to produce a number of gravitational lensing studies and publicly 
available products. Light cones of three different sizes (640, 340 and 123 deg$^2$) and depths have 
been constructed from the simulation snapshots and projected onto a series of shells. Ray-tracing through 
these cones using a modified version of the \textsc{glamer} \citep{2014MNRAS.445.1942M,2014MNRAS.445.1954P} 
code has been done to produce maps of shear, convergence and deflection for a range of source redshifts 
from 0 to 7 with one arcsecond resolution. For DR1, these maps are being made public along with accurate 
shear power spectra, correlation functions and the halo-shear correlations. In addition, source catalogs 
with random positions and known redshift distributions will be provided so that users can sample from them 
for Monte Carlo calculations.  In further data releases, with semi-analytic and Halo Occupation Distribution 
(HOD) models for the galaxies and inter-cluster baryons, we will produce galaxy-shear lensing predictions 
and strong lensing statistics for surveys such as Euclid. These results will be presented in a series 
of companion papers.

Mock catalogs constructed using three semi-analytic models, \nugc\ \citep{Makiya2016, Shirakata2019}, 
SAG \citep{Cora2018}, and SAGE \citep{Croton2016} will be released as DR2 in the near future.
All data will be made publicly available on \urluchuu. 

\section*{Acknowledgments}
The authors wish to acknowledge the referee of this paper, Chris Power, for his valuable comments.
We thank the Instituto de Astrof\'isica de Andaluc\'ia (IAA-CSIC), Centro de Supercomputaci\'n de Galicia (CESGA, \url{https://www.cesga.es}) and the Spanish academic and research network (RedIRIS, \url{http://www.rediris.es}) in Spain for hosting Uchuu DR1 in the \SU{} site (\urluchuu) for cosmological simulations. We are specially grateful to Antonio Fuentes, and his team at RedIRIS, for providing the skun@IAA\_RedIRIS server that hosts Uchuu DR1 through \SU{} with a fast download speed trough their RedIRIS High Speed Data Transfer Service; to Javier Cacheiro, Carlos Fern\'andez, Ignacio L\'opez, and Pablo Rey at CESGA for developing the Uchuu-BigData platform and providing the computer resources; and the staff at the IAA-CSIC computer department for proving support and managing the skun@IAA\_RedIRIS server.

We thank Kazuki Kinjo for helping to construct the Uchuu merger trees. We are grateful to our Uchuu Collaborators \'Angeles Molin\'e, Miguel S\'anchez-Conde and Alejandra Aguirre-Santaella for sharing preliminary results of their ongoing study on the properties of Uchuu subhalos, which served as a helpful sanity check to some of our results. Their subhalo work, that will include full details on Uchuu subhalo abundances, distributions and concentrations, will be published in a forthcoming publication. We also acknowledge Chi An Dong-P\'aez, IAA Summer-2020 student, for developing the Jupyter notebook \textit{Playing with Uchuu on your own computer}, and helping to test Spark notebooks on the Uchuu-BigData@CESGA. We would like to thank Britton Smith for quickly adding the capability to load the new merger tree format into \textsc{ytree}~\citep{ytree}.

The Uchuu simulations were carried out on Aterui II supercomputer at Center for Computational Astrophysics, CfCA, of National Astronomical Observatory of Japan, and the K computer at the RIKEN Advanced Institute for Computational Science (Proposal numbers hp180180, hp190161). The Uchuu DR1 effort has made use of the skun6@IAA facility managed by the IAA-CSIC in Spain, this equipment was funded by the Spanish MICINN EU-FEDER infrastructure grant EQC2018-004366-P. The skun@IAA\_RedIRIS server was funded by the MICINN grant AYA2014-60641-C2-1-P.
The numerical analysis were partially carried out on XC40 at the Yukawa Institute Computer Facility in Kyoto University.

TI has been supported by MEXT via the ``Priority Issue on Post-K computer''
(Elucidation of the Fundamental Laws and Evolution of the Universe), 
JICFUS, and MEXT via the``Program for Promoting Researches on the Supercomputer
Fugaku'' (Toward a unified view of the universe: from large scale
structures to planets, proposal numbers hp200124). 
TI thanks the support by MEXT/JSPS KAKENHI Grant Number
JP17H04828, JP18H04337, JP19KK0344, and JP20H05245. 
FP, AK and DM thank the support of the Spanish Ministry of Science and Innovation funding grant PGC2018-101931-B-I00.
Parts of this research were conducted under the Australian Research Council Centre of Excellence for All Sky Astrophysics in 3 Dimensions (ASTRO 3D), through project number CE170100013.
EJ acknowledges financial support from CNRS. 
FP and EJ want to thank a French-Spanish international collaboration grant from CNRS and CSIC.
CVM acknowledges support from FONDECYT through grant 3200918, and he also acknowledges previous support from the Max Planck Society through a Partner Group grant.
SAC acknowledges funding from
 Consejo Nacional de Investigaciones Cient\'{\i}ficas y T\'ecnicas (CONICET, PIP-0387), 
Agencia Nacional de Promoci\'on de la Investigaci\'on, el Desarrollo Tecnol\'ogico y la Innovaci\'on (Agencia I+D+i, PICT-2018-03743), and  Universidad Nacional de La Plata (G11-150), Argentina.

Numerical analysis were partially carried out using the following packages, \textsc{numpy} \citep{numpy}, 
\textsc{scipy} \citep{scipy}, \textsc{h5py}\citep{h5py},
\textsc{matplotlib} \citep{matplotlib}, 
\textsc{colossus} \citep{Diemer2018}, and \textsc{ytree} \citep{ytree}.
We thank the developer of \textsc{colossus}, Benedikt Diemer, 
for implementing the mass-concentration relation of this paper
in \textsc{colossus}. 

\section*{Data Availability}
As data release 1, 
we release various $N$-body products for the Uchuu suite on \urluchuu, 
such as subsets of simulation particles,
matter power spectra, halo/subhalo catalogues, and their merger trees. 
Non-public data are available upon request. 
Under the UchuuProject on Github~\footnote{\url{https://github.com/uchuuproject}}, 
we provide a collection of general tools for the community, 
codes / scripts / tutorials and Jupyter notebooks for the analysis of the Uchuu data.

\bibliographystyle{mnras}

\appendix

\section{How accurate are estimates of halo concentration?}\label{sec:m-c_accuracy}

Here we investigate the accuracy of our methods to estimate halo concentration.
We start with random realizations of dark matter halos having either NFW or Einasto density profiles. 
We then proceed by measuring concentrations using averaged halo profiles from the Uchuu 
simulation at $z=0$ and $z=1.4$. These results are compared between the Uchuu, Shin-Uchuu, 
and Phi-4096 simulations, each having significantly different mass resolution.

\makeatletter{}\begin{figure}
  \centering
\includegraphics[width=0.50\textwidth]{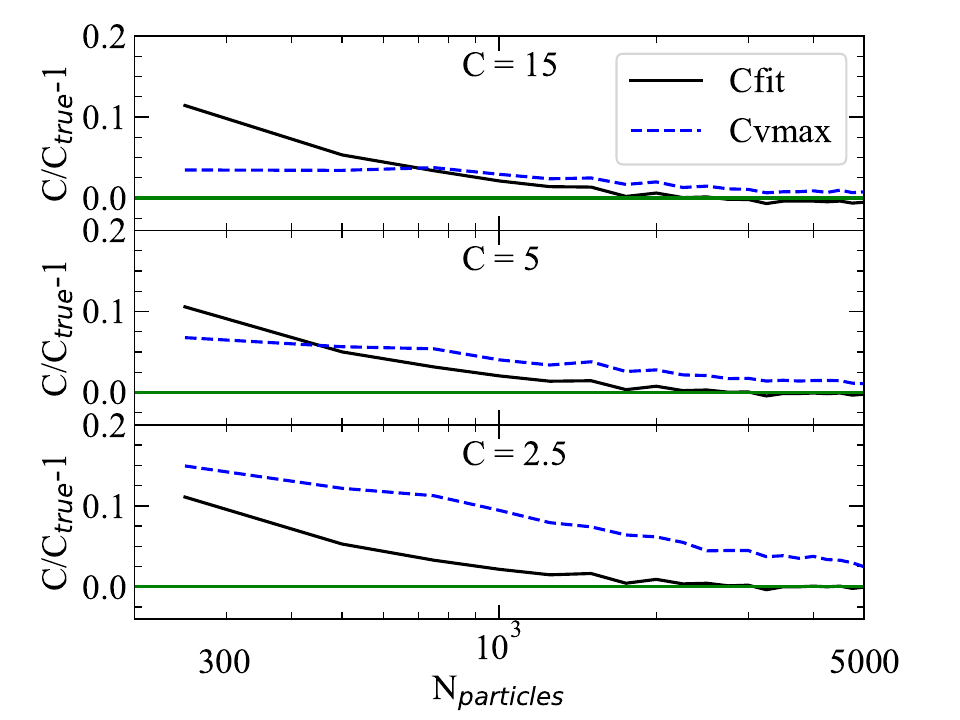}
\caption{Accuracy of recovering the concentration for halos with random
  realizations of the NFW density profile. The difference between the measured and true
  concentration is dependent on the number of
  halo particles $N_{\rm particles}$, the input concentration $C_{\rm true}$, and the
  measurement method. For very small number of particles
  ($N_{\rm particles}\lsim 500$), the $V_{\rm max}$ method provides more
  accurate results. For a large number of particles
  ($N_{\rm particles}\gsim 3000$), fitting the density profile gives
  better results. Overall the differences are small for concentration $C\gsim 3$,
  and less than $\sim 2$ per cent for $C_{\rm true}>5$ and $N_{\rm particles}>3000$.
  }
\label{fig:ANFW}
\end{figure}

\makeatletter{}\begin{figure}
  \centering
\includegraphics[width=0.45\textwidth]{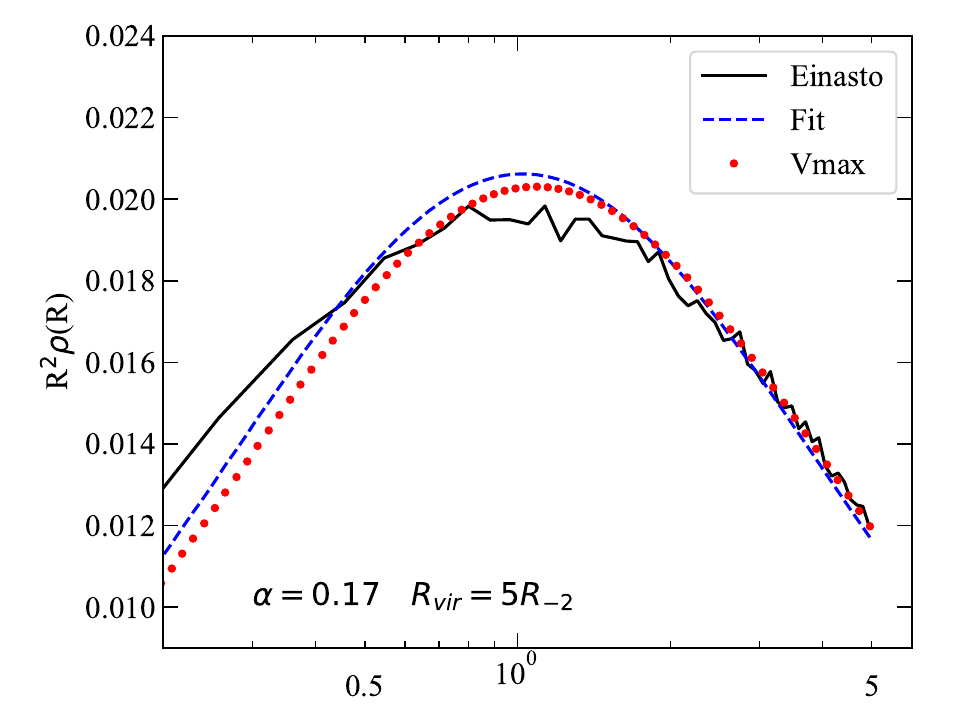}
\includegraphics[width=0.45\textwidth]{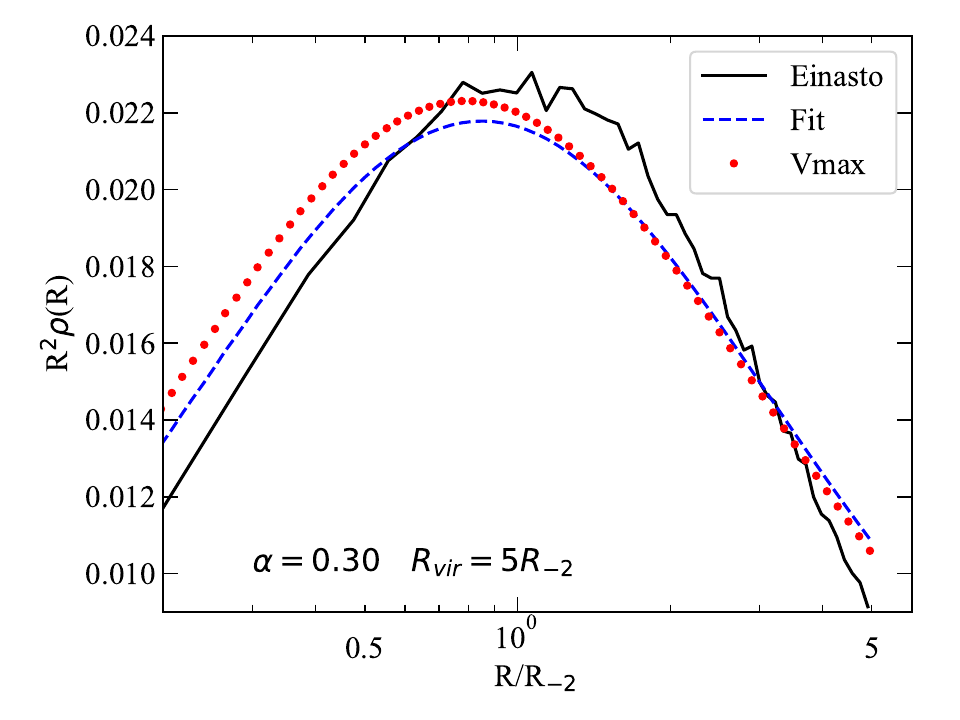}
\caption{Two examples of fitting Einasto profiles with the NFW profile and  
using two methods to measure the concentration. Parameters of the Einasto profile are 
indicated in the panels. For both cases a halo single realization with 300,000 particles was used. 
The profile in the top panel has a shape parameter $\alpha=0.17$ that can 
be reasonably well approximated by the NFW profile 
for radii $R\gsim 0.1R_{\rm vir}$. Indeed, the error in density at those 
distances are less than $\sim 5$ per cent. Both methods provide similar quality fits. 
The bottom panel shows the case of large shape parameter, $\alpha=0.3$, which is typical 
for very massive halos that form in high $\nu=\delta_{\rm c}/\sigma(M)\approx 4$ density peaks. 
The $V_{\rm max}$ method provides a better approximation for the outer region $R>0.1R_{\rm vir}$, 
while the direct fitting method is slightly better in the center. 
In both cases, the error on the density is $\sim 10$ per cent for 
$R>0.1R_{\rm vir}$. Density is shown in arbitrary units.
}
\label{fig:Einasto}
\end{figure}

\makeatletter{}\begin{figure}
  \centering
\includegraphics[width=0.5\textwidth]{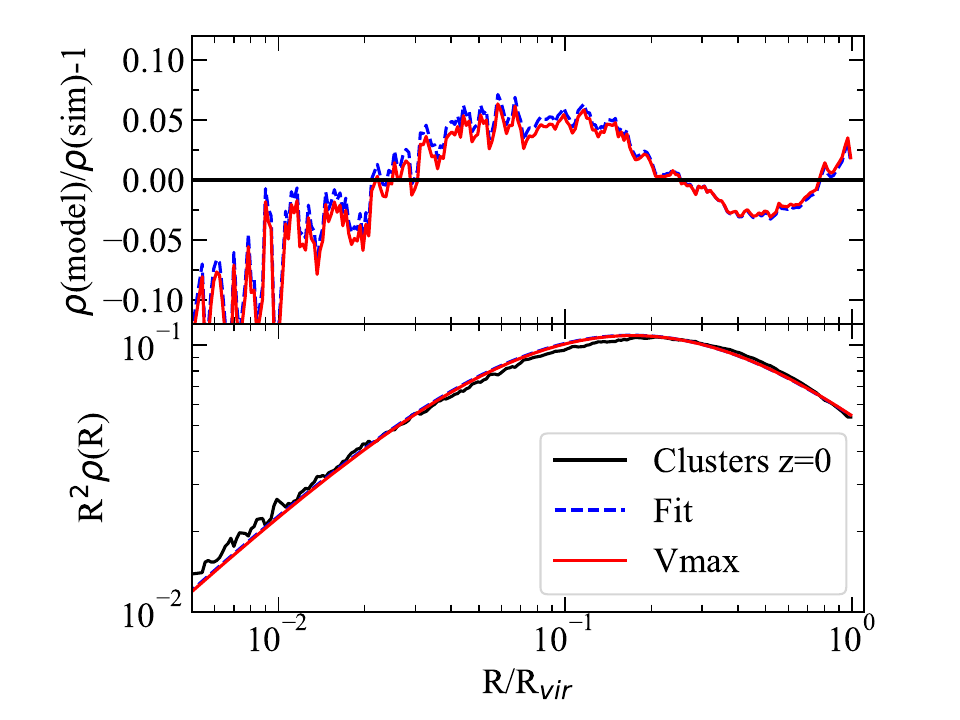}
\caption{Fitting the halo density profile of $M_{\rm vir}=1.5\times 10^{15}$\hMsun\
halos at $z=0$ with the NFW profile. The true halo profile was obtained 
by averaging the profiles of 100 halos of this mass in the Uchuu simulation. 
Both halo concentration measurement methods provide nearly identical fits, with 
a $\sim 5$ per cent accuracy for $R=(0.01-1)R_{\rm vir}$. 
The lower panel shows $R^2\rho(R)$ in arbitrary units.
}
\label{fig:ClustZ0}
\end{figure}

\subsection{Random realizations of NFW halos}
This is a simple test: make a random realization 
of a NFW halo with a given concentration and some number of particles. Because, by design, 
the concentration $C_{\rm true}$ is known, ideally the
algorithm should find it. However, there is always noise in any such measurement due to 
the finite number of particles, and the algorithm may produce biased estimates. 

We approximately follow the procedure used by the \textsc{rockstar} halo finder \citep{Behroozi2013}. 
All particles are sorted by distance. Binning is done by constant mass so that each interval 
has the same number of particles (S/N). The total number of bins is 50. However the first
bin is special: it always has 15 particles. We then measure the density and circular velocity 
for each bin. A density profile is then fitted using the NFW profile, providing a 
concentration value $C_{\rm fit}$. Similarly, a circular velocity curve is used to find 
$V_{\rm max}$, leading to an estimate of a concentration value $C_{\rm vmax}$. 
Many realizations (typically 300) are performed to find the average and rms measurement 
of each true concentration. 

Figure~\ref{fig:ANFW} present $C/C_{\rm true}$ as a function of particle number for each method. 
The three panels show three selected values for 
the true concentration: $C_{\rm true} =15$ for high 
concentration halos like our Milky Way, $C_{\rm true} =5$ for intermediate halos, 
and $C_{\rm true} =2.5$ for very low concentrations, close to the minimum concentration 
of the simulations at high redshifts.
The deviation from the true value is found to depend on the number of particles $N_{\rm particles}$, 
the input concentration $C_{\rm true}$, and the method used. For very small particle number 
($N_{\rm particles}\lsim 500$) the $V_{\rm max}$ method provides more accurate results. 
For large particle number ($N_{\rm particles}\gsim 3000$) fitting the density profile gives 
better results. However the differences are small for concentrations $C\gsim 3$, 
and less than $\sim 2$ per cent for $C_{\rm true}>5$ and $N_{\rm particles}>3000$. 
The only substantial errors happen at very low concentration where $C_{\rm max}$ 
can be systematically above the true values by 10-15 per cent. The main reason for this
is related to the fact that the radius corresponding to $V_{\rm max}$ is very close to 
the radius of the halo, and fluctuations are amplified producing an upward biased estimate.

The main take away of this test is that both methods produce nearly the same estimates
of the true concentration {\it if the density profile is NFW and the number of particles 
are more than 3000.}

\makeatletter{}\begin{figure}
  \centering
\includegraphics[width=0.5\textwidth]{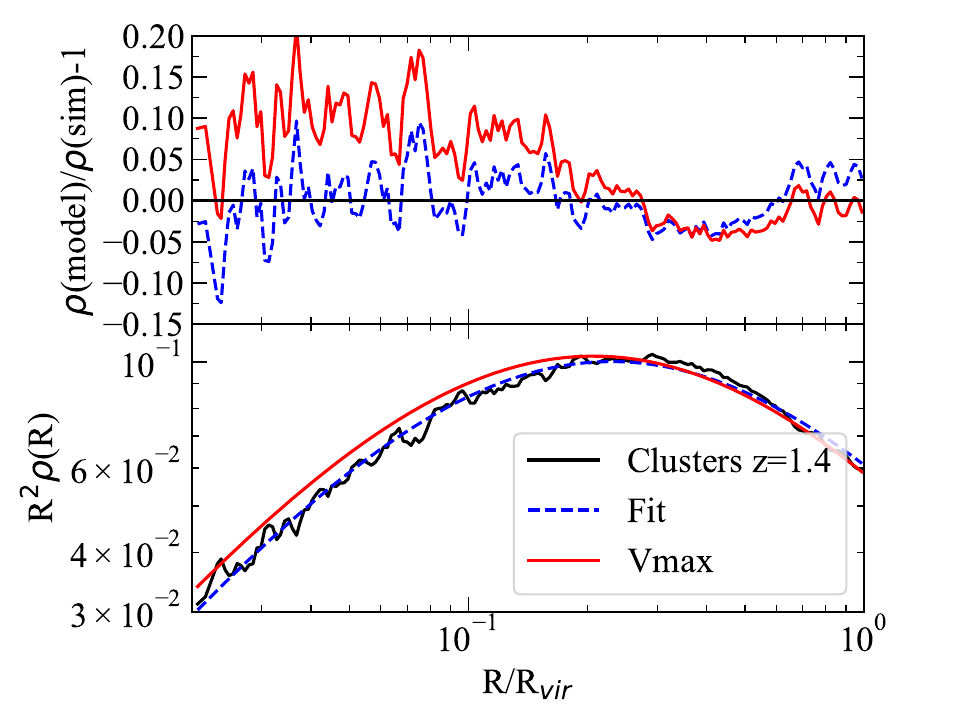}
\caption{Fitting the halo profile of $M_{\rm vir}=1.5\times 10^{14}$\hMsun\
halos at $z=1.4$ with the NFW profile. The true halo profile was obtained by 
averaging the profiles of 100 halos of this mass in the Uchuu simulation. 
Both halo concentration measurement methods provide nearly identical fits in
the outer halo region $R=(0.2-1)R_{\rm vir}$, accurate to $\sim 5$ per cent, 
while the fitting procedure does better in the center. The lower panel shows 
$R^2\rho(R)$ in arbitrary units.
}
\label{fig:ClustZ1.4}
\end{figure}

\begin{figure}
\centering 
\includegraphics[width=9cm]{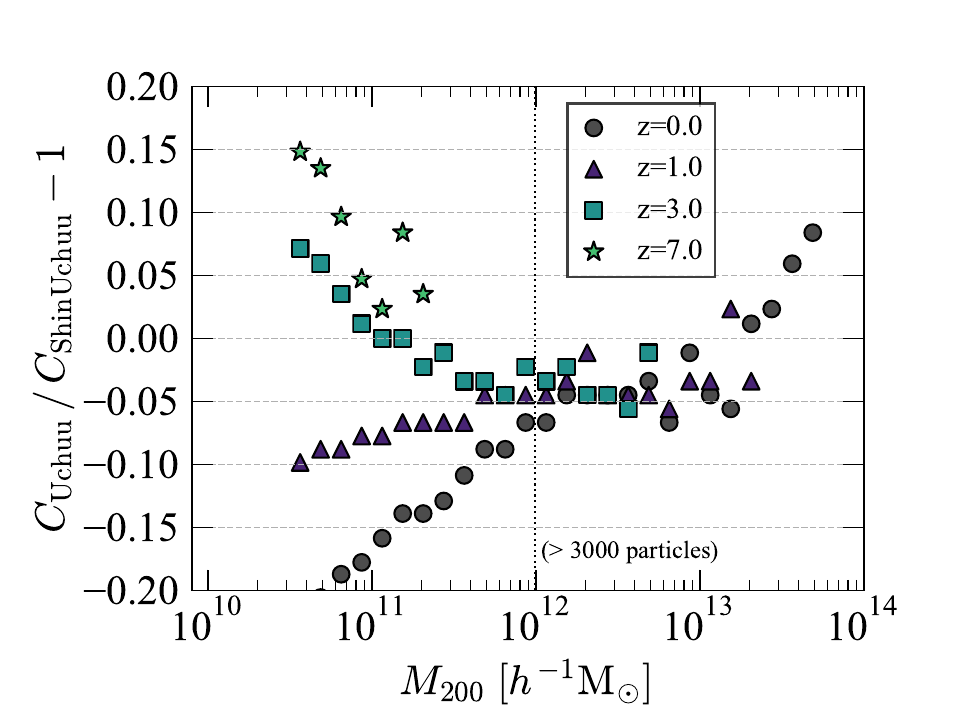}
\caption{
Relative residuals of the median values of
halo concentration $c_{200}$, comparing the Shin-Uchuu (higher resolution) and Uchuu 
(lower resolution) simulations, at $z=0.0, 1.0, 3.0$, and $7.0$. 
Concentrations are measured using the $V_{\rm max}$ method.
The vertical dotted line corresponds to a halo mass with 3,000 particles in the Uchuu simulation. 
At this point, residuals become larger with decreasing halo mass. 
}
\label{fig:m-c_res}
\end{figure}

\begin{figure}
\centering 
\includegraphics[width=9cm]{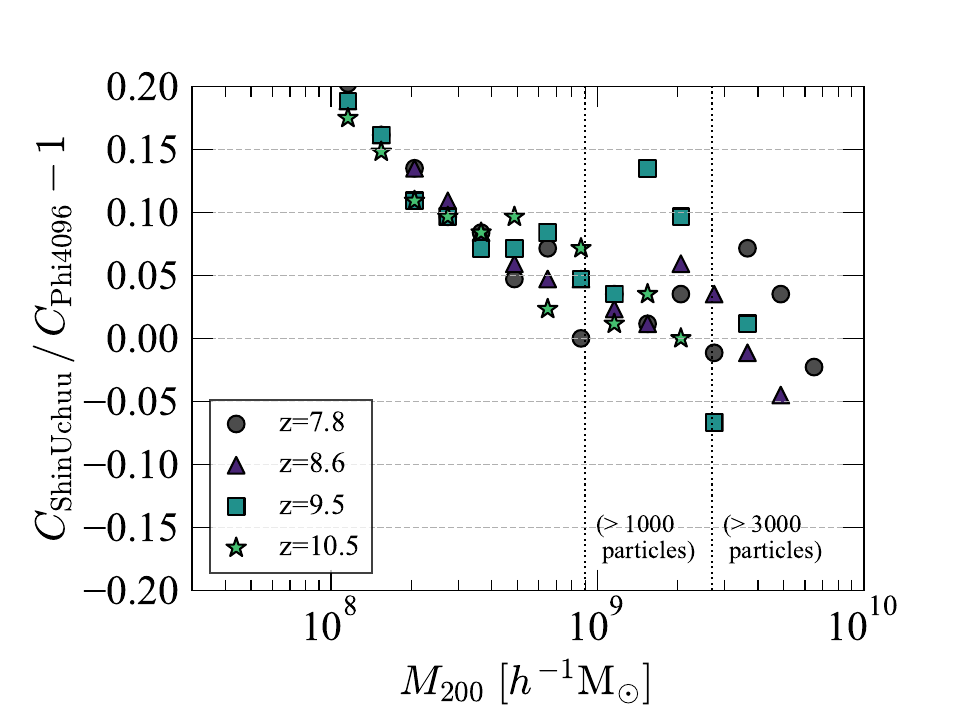}
\caption{
Relative residuals of the median values halo concentrations $c_{200}$, comparing the
Phi-4096 (higher resolution) and Shin-Uchuu (lower resolution) simulations, 
at $z=7.8, 8.6, 9.5$, and $10.5$. 
Concentrations are measured using the $V_{\rm max}$ method.
The vertical dotted lines corresponds to a halo mass with 1,000 and 3,000 particles 
in the Shin-Uchuu simulation. 
}
\label{fig:m-c_res2}
\end{figure}

\subsection{Random realizations of Einasto halos} 
Halos typically have profiles that are better approximated by the more general Einasto profile. 
The Einasto profile has three parameters, controlling the mass via $\rho_0$, 
the radius $R_{-2}$ where the log-log density slope is -2, and the shape parameter $\alpha$, 
which roughly describes the width of the profile. It has the functional form
\begin{equation}
   \rho(R)=\rho_0\exp\left[-\frac{2}{\alpha}\left(x^\alpha-1 \right)     \right], \quad x\equiv R/R_{-2}.
   \label{eq:Einasto}
\end{equation}

Figure~\ref{fig:Einasto} gives the example of two Einasto profiles with different shape parameters.
In the top panel we show $\alpha=0.17$, which is typical for galaxy-size halos. Such halos
have a relatively broad profile and are reasonably accurately fitted by the NFW profile. 
In the bottom panel we show $\alpha=0.30$, which is more characteristic of massive clusters or, 
in general, for high peaks in the initial density field. Because the shape of such halos are 
not NFW, there is no true concentration using the formal definitions. However, we  
can still apply our methods and see how accurately the density profile is described.

Not surprisingly, fits for $\alpha=0.17$, while not perfect, are relatively good. 
The errors where the density profile peaks (at 1/5 the halo radius)
are approximately 5 per cent and increase towards very small radii. 
The two methods of measuring concentration produce different estimates at the 
20 per cent level. However, it is not clear which one is better overall. 
We find the $V_{\rm max}$ method provides a slightly improved fit in the outer regions of the halo, 
while the fitting method is better towards the halo center.

The $\alpha=0.30$ halo is much more difficult to compare because the halo profile is far from NFW. 
The methods produce different concentrations, with $V_{\rm max}$ predicting higher concentration 
and smaller errors in the outer regions.

\subsection{Average profiles of massive halos in the Uchuu simulation} 
Here we extend the above analysis to real halos, and measure concentrations using the
averaged profiles of massive relaxed halos selected from a narrow mass interval.
We define a halo to be relaxed if its offset parameter (the distance between the halo center 
and halo center of mass, measured in units of the virial radius) is 
less than 0.05, and its spin parameter less than 0.03. This is done to remove halos that experience 
significant non-equilibrium events. 
Averaging over many halos of the same mass serves the same purpose as smoothing substructures.
Results are presented in Figures~\ref{fig:ClustZ0} and \ref{fig:ClustZ1.4} for $100$ 
clusters at $z=0$ and $z=1.4$. 

At $z=0$ both methods of measuring concentration produced very close estimates.
Note that the halo profile we find itself is not a NFW; there are $\sim 5$ per cent systematic deviations. 
The errors and the shape of the average halo profile is different at $z=1.4$ compared with $z=0$. 
Here, the fitting method works better except at the very outer 20 per cent region near the virial radius.
The difference in concentration produced by the two methods is about 10 per cent, 
with the $V_{\rm max}$ method predicting higher concentration. 

\subsection{The convergence of halo concentrations} 
Because we have three simulations with vastly different force and mass resolutions, 
we can compare estimates of halos concentration for halos of the same mass found in different simulations. 
We can also investigate the minimum number of particles per halo that is
required to determine the halo concentration robustly.

Figure~\ref{fig:m-c_res} shows the relative residuals of median
concentration between the Uchuu and the Shin-Uchuu simulations as a
function of halo mass.  The difference is less than around 5 per cent for
halos more massive than $\sim10^{12}$\hMsun, corresponding to
3,000 particles in the Uchuu simulation.  For less massive halos, the
difference increases regardless of the halo mass.  Therefore, we
conclude that 3,000 particles per halo is the minimum threshold for
the Uchuu simulation to robustly measure concentration.

Figure~\ref{fig:m-c_res2} shows the relative residuals of median
concentration between the Shin-Uchuu and the Phi-4096 simulations as
the function of halo mass.  Differences can be seen to increase systematically
below $\sim 9.0 \times 10^{8}$\hMsun, corresponding to 1,000
particles in the Shin-Uchuu simulation.  For more massive
halos the difference is less than 5 per cent, except at $z=9.5$ 
where the scatter is large due to poor
statistics in the Phi-4096 simulation. However, we observe no systematic
trend. Therefore, we consider 1,000 particles per halo is the minimum
threshold for the Shin-Uchuu simulation to robustly measure concentration.  
This value is smaller than
that for the Uchuu simulation, reflecting that the softening length of Shin-Uchuu
is much smaller.  We also use this threshold for the
Phi-4096 simulation because the ratio of mean particle separation to
softening length is close to that of the Shin-Uchuu simulation.


\section{Analytical model of mass-concentration relation}\label{sec:m-c_model}

Following \citet{Diemer2019}, we model halo concentration using the analytical approximation
\begin{equation}
c(\nu,n_{\rm eff},\alpha_{\rm eff}) = C\left(\alpha_{\mathrm{eff}}\right) \times 
\tilde{G}\left(\frac{A\left(n_{\mathrm{eff}}\right)}{\nu}\left[1+\frac{\nu^{2}}{B\left(n_{\mathrm{eff}}\right)}\right]\right), \label{eq:conc}
\end{equation}
where $\tilde{G}(x)$ is the inverse function of
\begin{equation}
G(x)=\frac{x}{[f(x)]^{\left(5+n_{\mathrm{eff}}\right) / 6}}. 
\end{equation}
Here, $f(x) =\ln(1+x)-x/(1+x)$ is the mass function of the NFW profile, 
$\nu =\delta_{\rm c}/\sigma(M)$ is the height of the density peak, 
$\delta_{\rm c}=1.686$ is the critical over-density for spherical collapse,  
and $\sigma(M)$ is the rms density fluctuation. 
Variables $n_{\rm eff}$ and $\alpha_{\rm eff}$ are defined as
\begin{equation}
n_{\mathrm{eff}}(M)=-\left.2 \frac{d \ln \sigma(R)}{d \ln R}\right|_{R=\kappa R_{\mathrm{L}}}-3, \label{eq:m-c2}
\end{equation}
and
\begin{equation}
\alpha_{\mathrm{eff}}(z)=-\frac{d \ln D(z)}{d \ln (1+z)}.
\end{equation}
The latter is the effective exponent of linear growth $D(z)$. 
The former reflects the effective slope of the power spectrum,
$\sigma(R)$ is the rms density fluctuation 
in spheres with Lagrangian radius $R_{\rm L}$ multiplied by a free parameter $\kappa$. 
Halo mass $M$ is directly related to $R_{\rm L}$ through
\begin{equation}
M = \frac{4\pi}{3} \rho_m R_{\rm L}^3,
\end{equation}
where $\rho_m(z=0)$ is the mean density at $z=0$.
Terms
$A(n_{\rm eff})$, $B(n_{\rm eff})$, and $C(\alpha_{\rm eff})$ in Equation~\ref{eq:conc} have the following form:
\begin{equation}
\begin{array}{l}
A\left(n_{\mathrm{eff}}\right)=a_{0}\left(1+a_{1}\left(n_{\mathrm{eff}}+3\right)\right), \\
B\left(n_{\mathrm{eff}}\right)=b_{0}\left(1+b_{1}\left(n_{\mathrm{eff}}+3\right)\right), \\
C\left(\alpha_{\mathrm{eff}}\right)=1-c_{\alpha}\left(1-\alpha_{\mathrm{eff}}\right),  \label{eq:m-c1}
\end{array}
\end{equation}
with free parameters $a_0, a_1, b_0, b_1$, and $c_{\alpha}$.

\section{Mass-concentration relation for relaxed halos and for overdensity 500 halo definition}\label{sec:m-c_relaxed}
\begin{figure}
\centering 
\includegraphics[width=9cm]{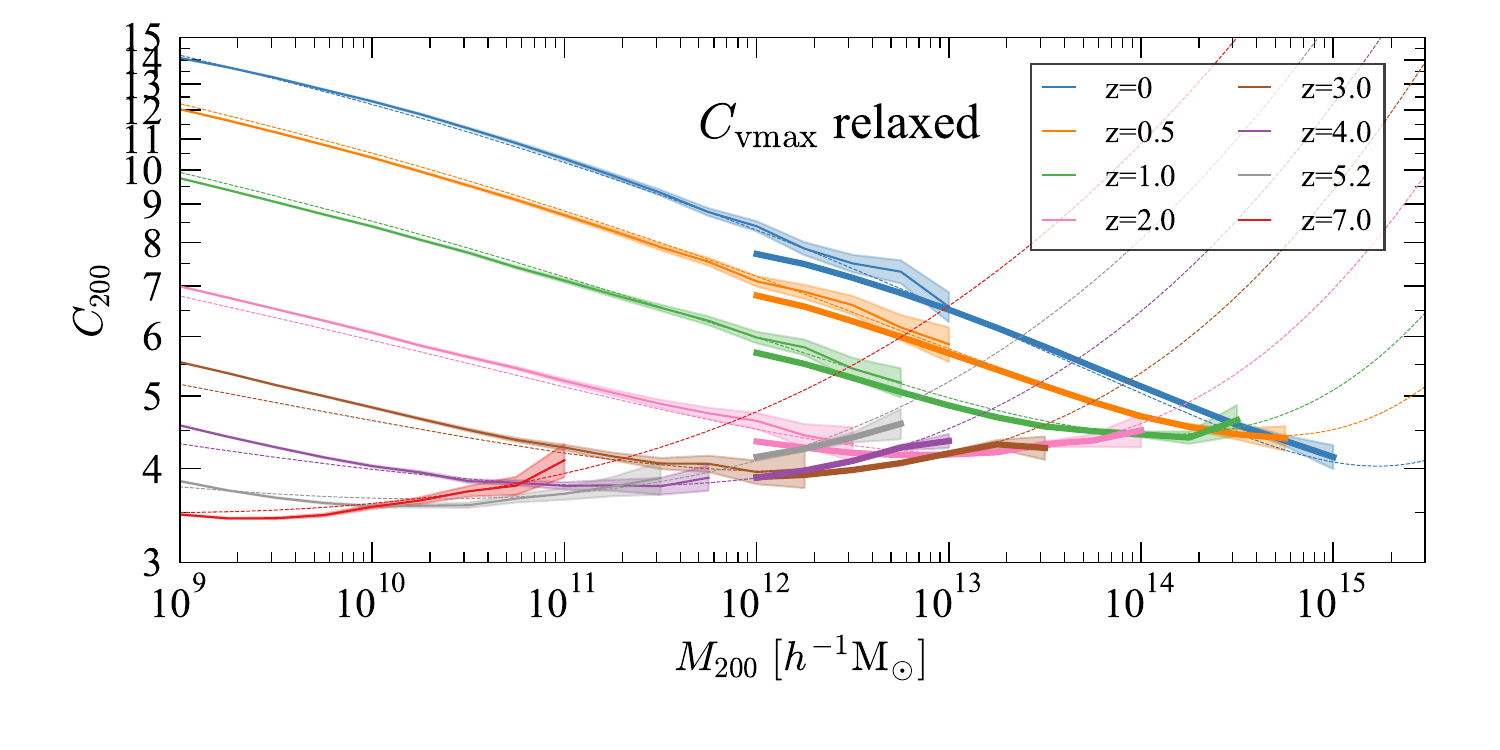} 
\includegraphics[width=9cm]{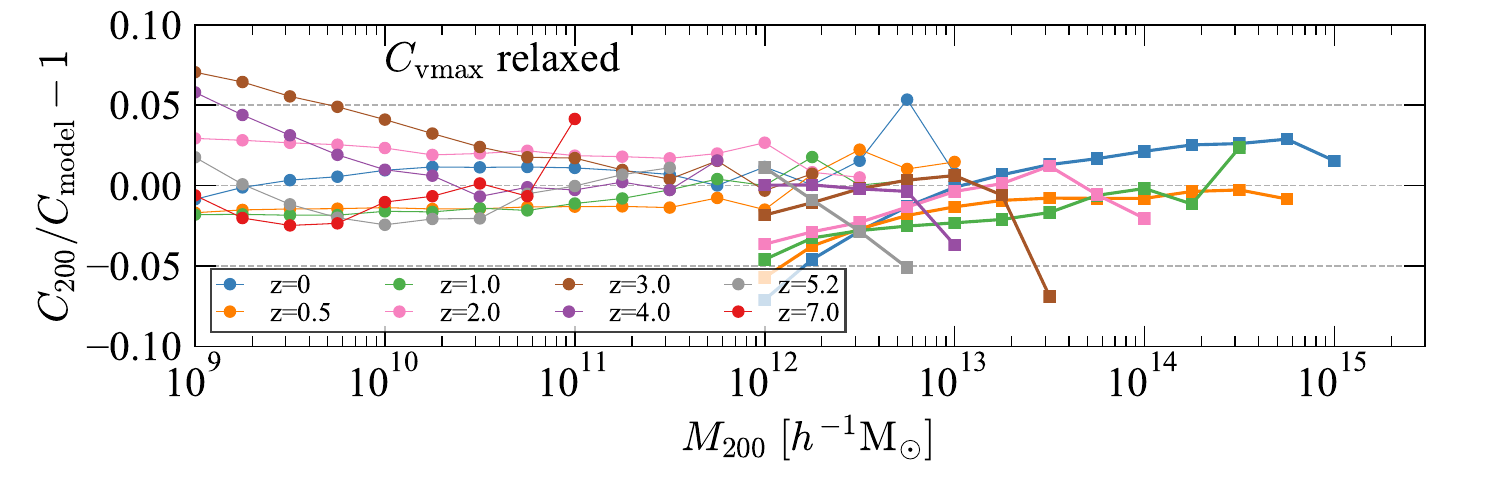} 
\caption{
Concentrations for relaxed halos using $V_{\rm max}$ method.
(eq.(\ref{eq:v200})) 
Top panel: Mass-concentration relation of relaxed halos for the Uchuu (thick curves) and the Shin-Uchuu (thin curves) simulations.  The dashed curves are predictions of the analytical model described by eq. (2). Bottom panel: The fractional difference between the halo concentrations of the Uchuu (squares) and the Shin-Uchuu (circles) simulations and the model predictions. 
}
\label{fig:m-c_relaxed}
\end{figure}

\begin{figure}
  \centering
\includegraphics[width=9cm]{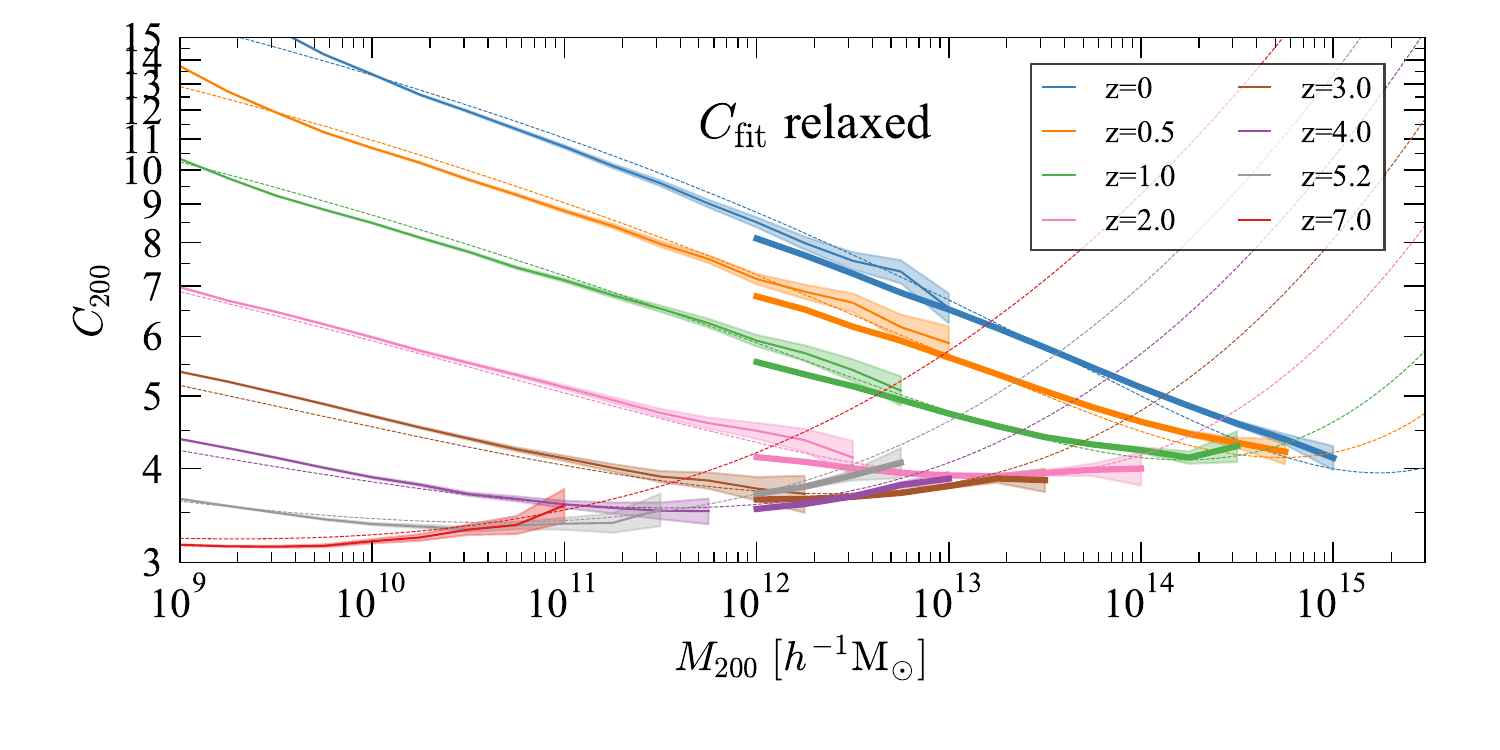} 
\includegraphics[width=9cm]{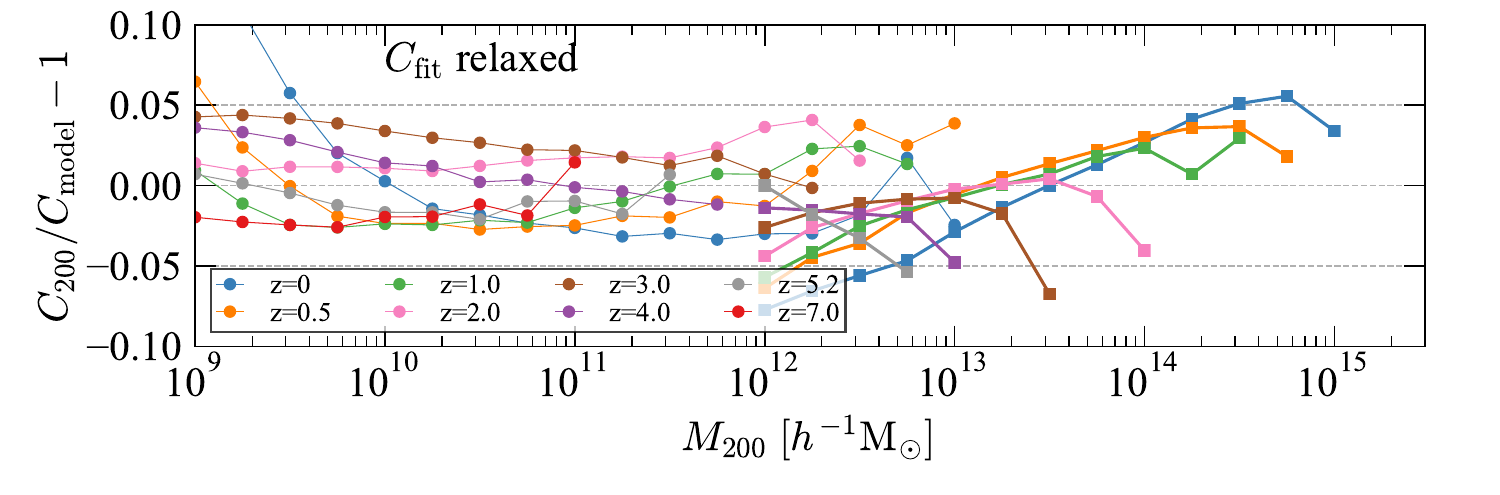} 
\caption{
The same as in Figure~\ref{fig:m-c_relaxed} but for concentrations estimated by profile fitting.
}
\label{fig:m-c_rs_relaxed}
\end{figure}

Figure~\ref{fig:m-c_relaxed} 
shows the mass-concentration relation of the
Uchuu and Shin-Uchuu simulations and their redshift evolution for only relaxed halos.
The concentrations are estimated using the $V_{\rm max}$ method (eq.(\ref{eq:v200})).
Overall trend is the same as the relation for all halos (Figure~\ref{fig:m-c}).
The concentration for relaxed halos shows a well-known behaviour
that relaxed halos have larger concentrations.
Even for relaxed halos, we observe a flattening and an upturn with
increasing mass, consistent with \citet{Klypin2016}. 
Figure~\ref{fig:m-c_rs_relaxed} also shows the mass-concentration relation, 
where concentrations are estimated by the profile fitting. 
The plot shows that the main tendencies of the concentration-mass
relations do not depend on the method to find concentration. 
This includes the upturn in the concentration at large masses.

\begin{figure}
\centering 
\includegraphics[width=9cm]{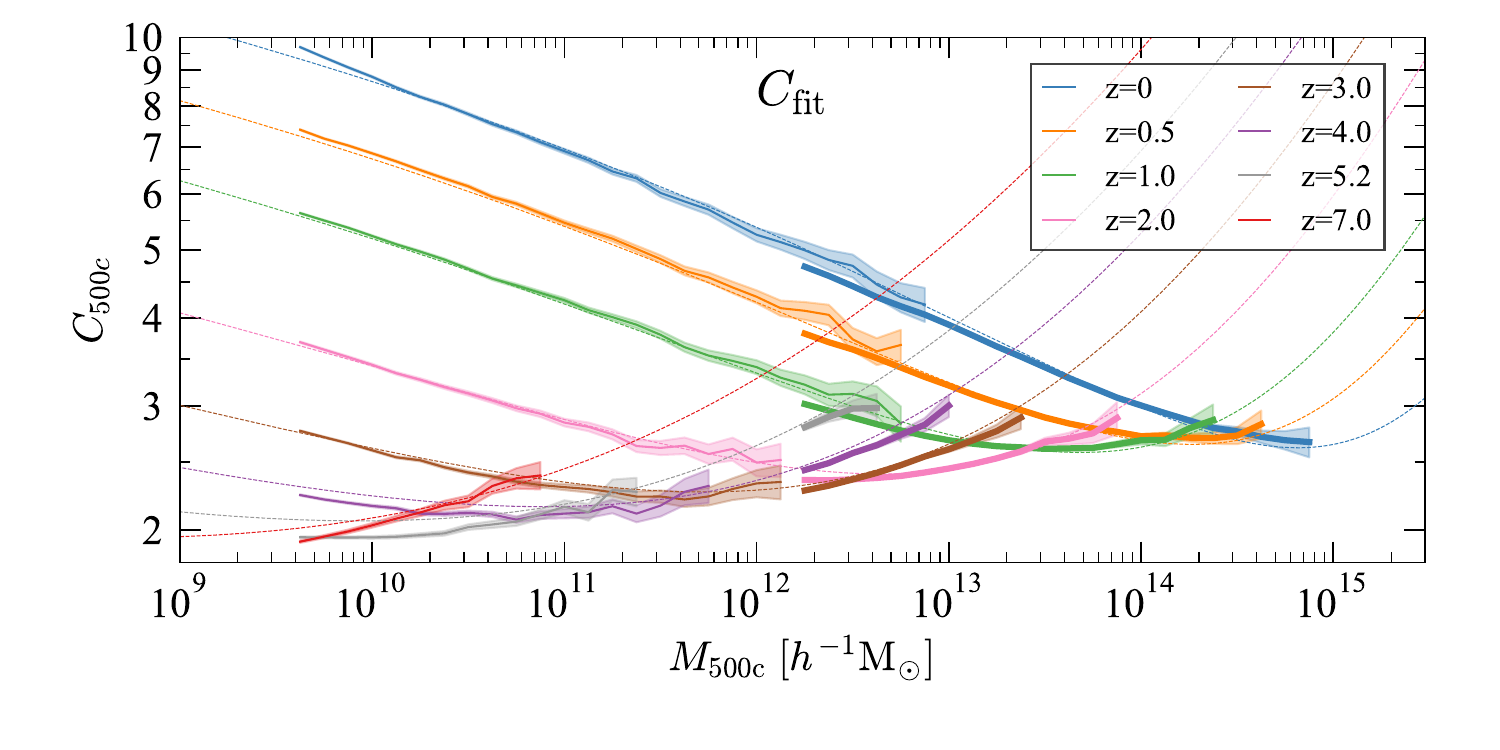} 
\includegraphics[width=9cm]{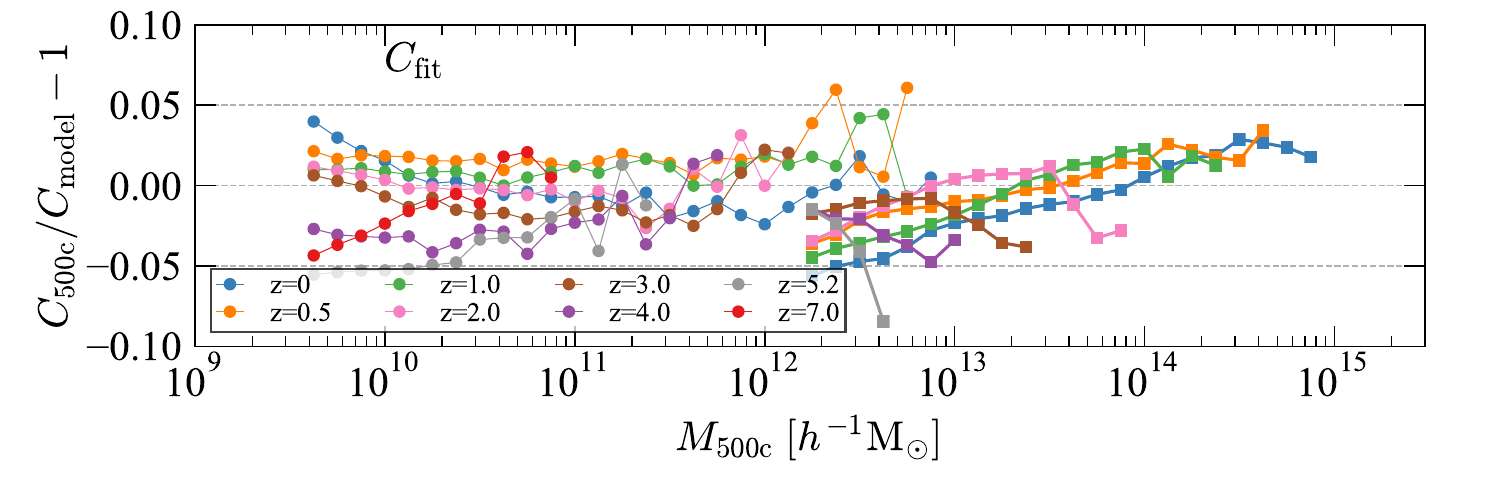} 
\caption{
Evolution of concentration-mass relation for all halos. Concentrations are estimated by profile fitting
and for  the halo mass and concentration  defined using $r_{500}$,
in which the average spherical overdensity is 500 times the critical density.
Labeling of curves is the same as in Figure~\ref{fig:m-c_relaxed}.
}
\label{fig:m-c500}
\end{figure}

Bottom panels of Figure~\ref{fig:m-c_relaxed} and \ref{fig:m-c_rs_relaxed} show
residuals of concentrations from the model.
The accuracy level is similar to those for all halos
(Figure~\ref{fig:m-c} and \ref{fig:m-c_rs}). 
We also confirm that the accuracy level is also comparable 
at mass $M_{200} < 10^{11}$\hMsun\ and redshift $z>7$.
We also provide best fitting parameters of the mass-concentration relation with $c_{500}$ 
estimated by profile fitting in Table~\ref{tab:fit_param} and show the result in Figure~\ref{fig:m-c500}.

\section{Validation of merger trees}\label{sec:merger_tree}

Our use of \textsc{consistent trees} needed to be modified to make
merger tree construction possible on simulations of the size of those
in the Uchuu suite.  As explained in \S \ref{sec:method}, to overcome
this hurdle, we split the full box of each simulation into smaller
regularly spaced sub-volumes, and ran \textsc{consistent trees} for
each sub-volume separately.  In this Section, we compare merger trees
of the mini-Uchuu simulation constructed by this method and the original
\textsc{consistent trees} code in order to show the robustness of this
method.

A number of metrics have been used to evaluate merger trees 
such as length of tree and mass growth history of halos 
\citep[e.g.,][]{Srisawat2013}, and merger rate. 
Visualization tools have also been provided 
\citep[e.g.,][]{ytree, Poulton2018}. 
Here, we use mean merger rate because mergers trigger a number of events 
in semi-analytic galaxy models. 
Given a halo with mass $M_i$ at redshift $z_i$, 
we define the merger mass ratio $\xi = M_{i-1,j} / M_{i-1,0}$, 
where $i$ and $j$ denote the snapshot id and progenitors of the halo, respectively. 
We set the progenitor with $j=0$ to be the most massive progenitor
and compute the number of mergers per halo as a function of $z_i$, $\xi$ 
and the descendant halo mass $M_i$. 

\begin{figure}
\centering 
\includegraphics[width=8.8cm]{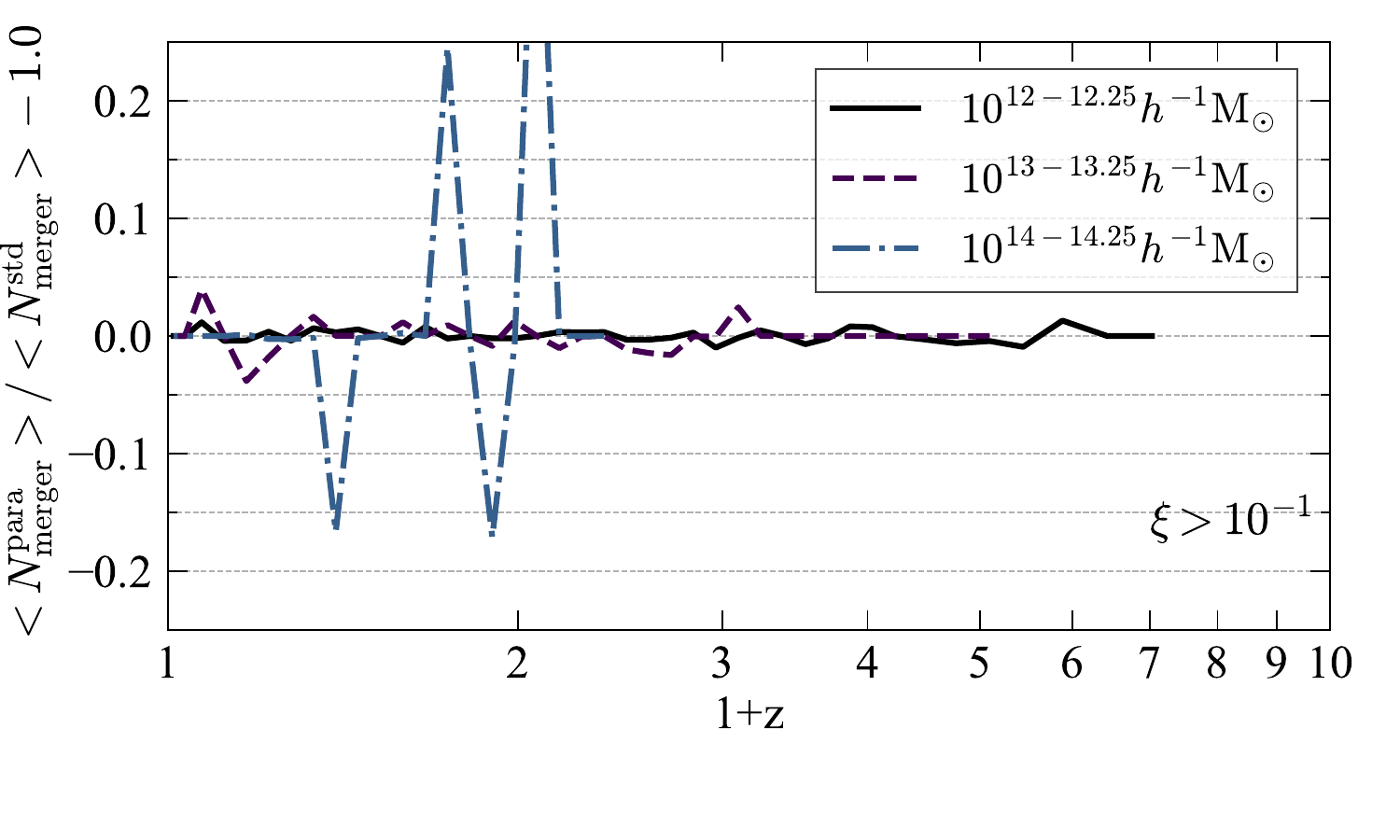} 
\includegraphics[width=8.8cm]{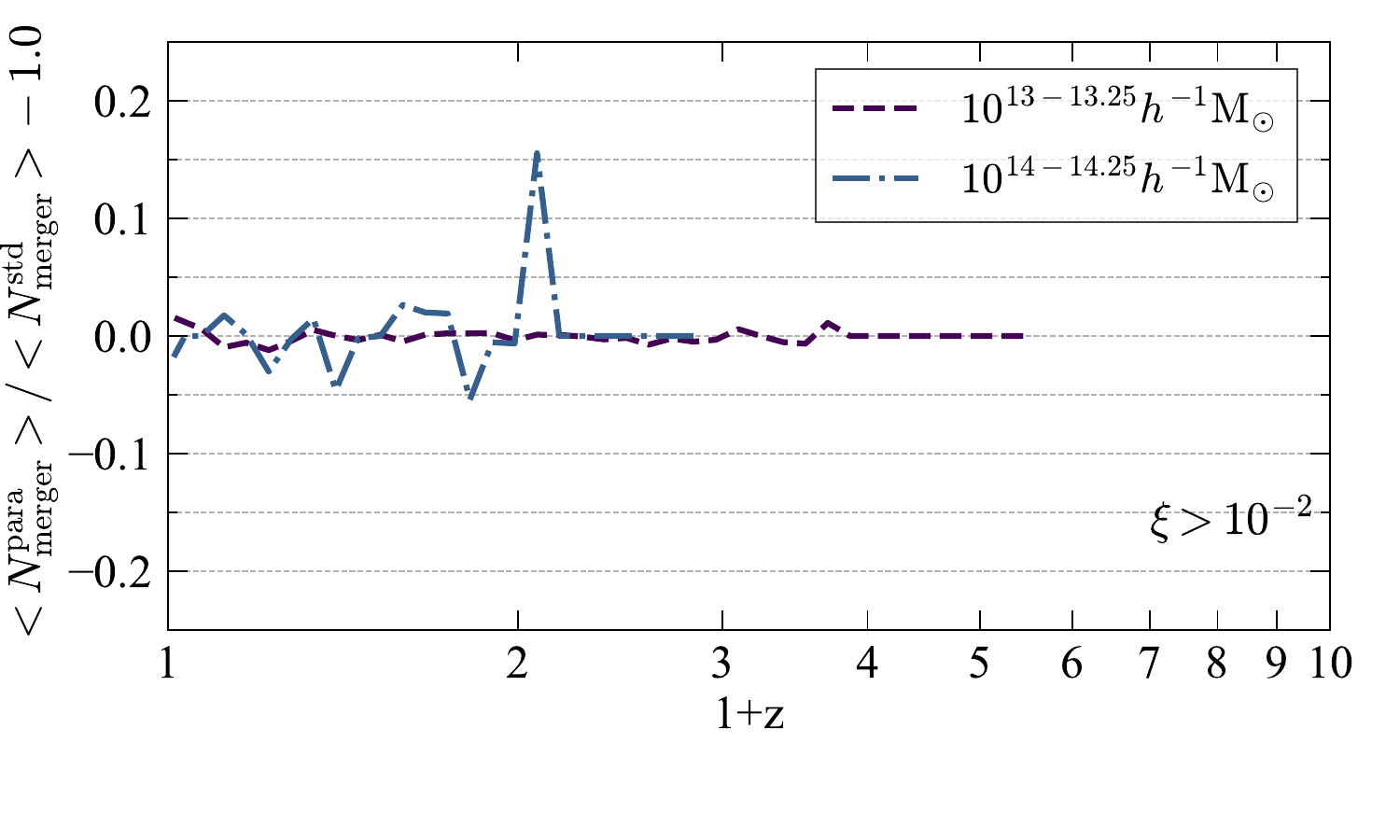}
\includegraphics[width=8.8cm]{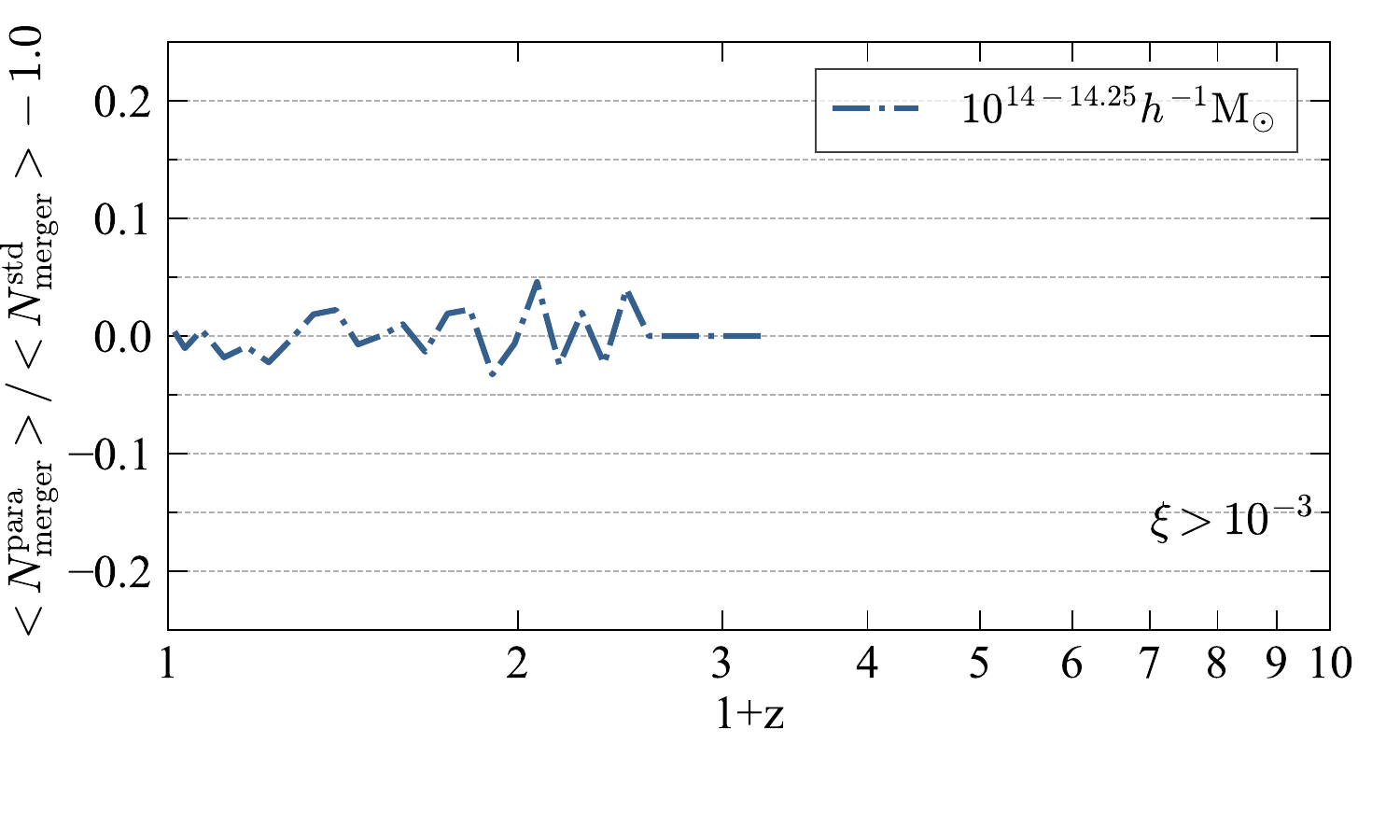}
\caption{
Comparison of mean merger rate
of trees of the mini-Uchuu simulation constructed by 
our method ($\langle N^{\rm para}_{\rm merger} \rangle$) described in \S~\ref{sec:method} and 
the original \textsc{consistent trees} code ($\langle N^{\rm std}_{\rm merger} \rangle$)
as a function of redshift for three halo mass ranges, and three merger mass ratios $\xi$. 
}
\label{fig:mrate}
\end{figure}

\begin{figure}
\centering 
\includegraphics[width=8.8cm]{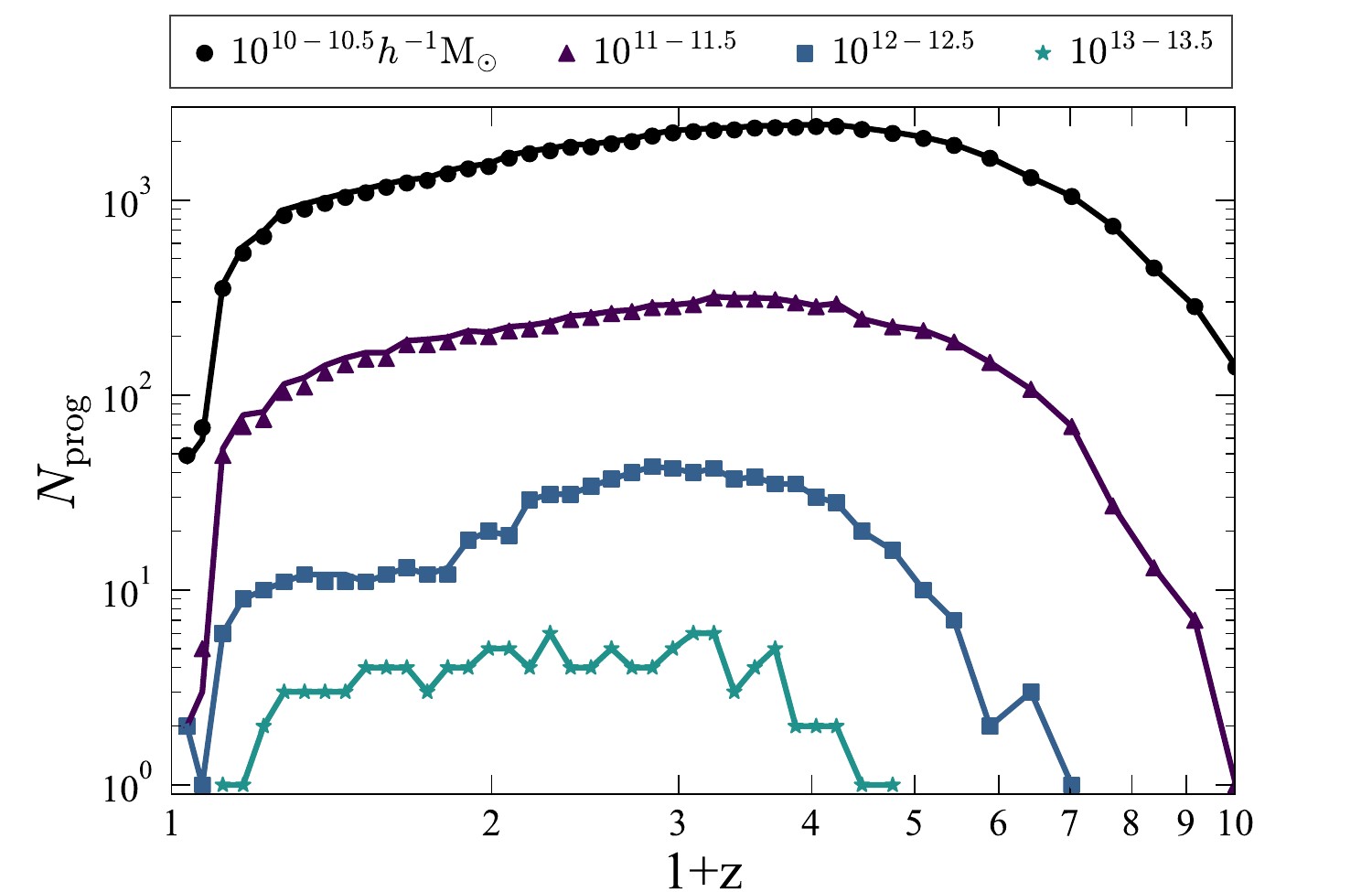} 
\caption{
Total number of progenitor halos and subhalos $N_{\rm prog}$ of the tree of the largest
halo at $z=0$ in the mini-Uchuu simulation as a function of redshift
for four progenitor mass ranges. Symbols denote results of our
method, and solid curves denote those of the original
\textsc{consistent trees} code.
}
\label{fig:mrate2}
\end{figure}

Figure \ref{fig:mrate} shows the comparison of mean merger rate
of trees of the mini-Uchuu simulation constructed by 
our method ($\langle N^{\rm para}_{\rm merger} \rangle$) and 
the original \textsc{consistent trees} code ($\langle N^{\rm std}_{\rm merger} \rangle$)
as a function of redshift for three halo mass bin 
and three merger mass ratio. 
Except for the most massive halo mass ($10^{14-14.25}$\hMsun), 
the difference of both methods is always within five per cent, 
ensuring the accuracy of our method. 
For halos with $10^{14-14.25}$\hMsun, both merger rates 
converge well within five per cent level with little exceptions. 
There are four spots for $\xi>10^{-1}$ and 
one spot for $\xi>10^{-2}$, where the difference exceeds ten per cent.
This is because such mergers with massive halos 
are rare and the number difference at each redshift is only one. 
Thus, this large difference in ratio is not significant. 

We also perform a stricter test.  Figure \ref{fig:mrate2} shows the
comparison of the total number of progenitor halos and subhalos of the
tree of the largest halo ($\sim 2.0 \times 10^{15}$\hMsun)
at $z=0$ in the mini-Uchuu simulation as a
function of redshift for four progenitor mass ranges.
The number of progenitors agrees well with each other
for progenitor mass more massive than $10^{12}$\hMsun.
For less massive progenitors there are small deviations,
but which are typically only less than five percent level.
We conclude that merger trees constructed by our method are equivalent to 
those obtained from the original \textsc{consistent trees} code.

\bsp	
\label{lastpage}
\end{document}